\documentclass[a4paper,11pt]{article}
\pdfoutput=1 
\usepackage{jheppub} 
\usepackage[utf8]{inputenc}
\usepackage{amsthm}
\usepackage{mathrsfs} 
\usepackage[all]{xy}
\graphicspath{{./images/}}
\theoremstyle{plain}

\theoremstyle{definition}

\numberwithin{thm}{section}

\newcommand{\gsim}{ \mathop{}_{\textstyle \sim}^{\textstyle >} }
\newcommand{\lsim}{ \mathop{}_{\textstyle \sim}^{\textstyle <} }
\newcommand{\vev}[1]{ \left\langle {#1} \right\rangle }

\def\diag{\mathop{\rm diag}\nolimits}

\DeclareMathOperator{\Tr}{Tr}
\DeclareMathOperator{\tr}{tr}

\def\Im{\mathop{\mathrm{Im}}}
\usepackage{slashed}

\def\CA{{\cal A}}

\def\CF{{\cal F}}

\def\CL{{\cal L}}

\def\CN{{\cal N}}

\def\BC{{\mathbb C}}

\def\BR{{\mathbb R}}

\def\BZ{{\mathbb Z}}

\def\sC{{\mathscr C}}
\def\sD{{\mathscr D}}
\def\sR{{\mathsf R}}
\def\sT{{\mathsf T}}

\def\U{\mathrm{U}}
\def\SU{\mathrm{SU}}
\def\O{\mathrm{O}}
\def\SO{\mathrm{SO}}
\def\Sp{\mathrm{Sp}}

\def\Spin{\mathrm{Spin}}
\def\Pin{\mathrm{Pin}}

\def\beq#1\eeq{\begin{align}#1\end{align}}

\title{Anomaly matching in QCD thermal phase transition 
}

\preprint{}

\author{Kazuya Yonekura}
\affiliation{Faculty of Arts and Science, Kyushu University, 
 Fukuoka, Fukuoka, 819-0395, Japan
}

\abstract{
We study an 't~Hooft anomaly of massless QCD at finite temperature.
With the imaginary baryon chemical potential at the Roberge-Weiss point, there is a $\mathbb{Z}_2$ symmetry which can be used to define confinement.
We show the existence of a mixed anomaly between the $\mathbb{Z}_2$ symmetry and the chiral symmetry, which gives a strong relation between confinement and chiral symmetry breaking.
The anomaly is a parity anomaly in the QCD Lagrangian reduced to three dimensions.
It is reproduced in the chiral Lagrangian by a topological term related to Skyrmion charge, matching the anomaly before and after QCD phase transition.
The effect of the imaginary chemical potential is suppresssed in the large $N$ expansion, and we discuss implications
of the 't~Hooft anomaly matching for the nature of QCD phase transition with and without the imaginary chemical potential. Arguments based on universality alone are disfavored, and
a first order phase transition may be the simplest possibility if the large $N$ expansion is qualitatively good.
} 

\begin{document}

\maketitle

\section{Introduction and summary} \label{sec:intro}

QCD phase transition is a very important problem in high energy physics and cosmology, and
it is also an extremely difficult problem. Until recently, there were almost no rigorous results about the nature of QCD phase transition.
There have been mainly two approaches to the problem. One approach is numerical lattice simulation, and the other is by assuming effective theories of Landau-Ginzburg type based on the argument of universality.

At the point of physical quark masses,
what is often said is that the QCD phase transition is cross-over, i.e., there is no definite phase transition and thermodynamic quantities behave smoothly 
as the temperature is changed. This is suggested by lattice simulations~\cite{Aoki:2006we,Bhattacharya:2014ara}. It is also suggested by the argument of universality in chiral symmetry breaking
$\SU(2)_L \times \SU(2)_R \to \SU(2)$ if the up and down quark masses $m_{u,d}$ are regarded as small (but nonzero) and the strange quark mass $m_s$ as heavy~\cite{Pisarski:1983ms}.
In that case, the phase transition may be second order in the limit $m_{u,d} \to 0$ if we assume universality,\footnote{
However, there is also a possibility that the anomalous axial symmetry $\U(1)_A$ is recovered to a very good approximation at finite temperature.
If that happens, even the conclusion based on universality can change. See e.g. \cite{Aoki:2012yj,Fukaya:2017wfq,Chiu:2013wwa,Nakayama:2014sba} for some recent studies and references therein. }
and the small nonzero $m_{u,d}$ change the transition from second order to cross-over.

However, because of difficulties of numerical lattice simulation in small quark mass region,
it is important to study the overall picture rather than just specific quark masses, and perform consistency checks to really firmly establish such results. For example, we can vary
quark masses to study the so-called Columbia plot of QCD phase diagram, and we can also vary the baryon chemical potential as a continuous parameter.  
See e.g. \cite{Fukushima:2010bq} for a review. 

In fact, some of the results obtained so far are still not consistent with each other. 
For example, it is not yet clear whether the phase transition is first order or second order in the chiral limit $m_{u,d} \to 0$ or $m_{u,d,s} \to 0$, and different studies give different results.
For brief summaries of the current situation, see e.g. \cite{Ding:2017giu,Aoki-san}.

Because of the above situation, it is important to find rigorous results which do not rely on numerical simulation or the assumption of universality.
It was difficult to obtain such rigorous results in finite temperature cases. However, a great progress was made towards this direction in \cite{Gaiotto:2017yup}.\footnote{
For earlier attempts, see \cite{Itoyama:1982up} in which the usual perturbative anomaly was considered rather than global anomalies. }
In that work, pure Yang-Mills theories with the topological $\theta$ angle at $\theta = \pi$ have been studied at finite temperature, by using 't~Hooft anomalies.
By an 't~Hooft anomaly, we mean an anomaly of global symmetries which exists if the global symmetries are gauged, as in the 't~Hooft's consideration of chiral symmetry breaking
by using the anomaly of the chiral symmetry $\SU(N_f)_L \times \SU(N_f)_R$ (see \cite{Weinberg:1996kr} for a standard textbook).
In \cite{Gaiotto:2017yup}, a very subtle 't~Hooft anomaly was found which survives even in finite temperature case, and it is used to severely constrain
the nature of the phase transition in pure Yang-Mills theories. Such subtle 't~Hooft anomalies are very useful for 
four dimensional gauge theories~\cite{Tachikawa:2016xvs,Yamazaki:2017ulc,Tanizaki:2017bam,Komargodski:2017smk,Shimizu:2017asf,Kikuchi:2017pcp,Gaiotto:2017tne,
Poppitz:2017ivi,DiVecchia:2017xpu,Kitano:2017jng,Tanizaki:2017qhf,Yamazaki:2017dra,Tanizaki:2017mtm,Cherman:2017dwt,Guo:2017xex,Draper:2018mpj,Seiberg:2018ntt,Shifman:2018yxh,
 Ritz:2018mce,Aitken:2018kky,Aitken:2018mbb,Anber:2018iof,Argurio:2018uup,Cordova:2018acb,Anber:2018jdf,Tanizaki:2018wtg,Bi:2018xvr,Yamaguchi:2018xse,Anber:2018xek,Bashmakov:2018ghn,
 Hsin:2018vcg,Wan:2018zql} as well as lower dimensional strongly coupled systems.

For the applications to QCD phase transition, the most relevant anomaly found so far is the one discussed in \cite{Shimizu:2017asf} (see also \cite{Tanizaki:2017qhf,Tanizaki:2017mtm}).
In the present paper,
we further study this direction (but the present paper is more elementary and self-contained). 
In \cite{Shimizu:2017asf}, a subtle anomaly has been found when there is an 
imaginary baryon chemical potential $\mu_B$ at a special value $\mu_B=\pi$,\footnote{In \cite{Shimizu:2017asf}, also a speculative discussion was given about the case of zero chemical potential. 
In any case, the effect of the imaginary chemical potential is sub-leading in the large $N$ expansion as we discuss later. }
where $\mu_B$ is normalized to be dimensionless.
Finite temperature QCD with imaginary chemical potential is a very important subject, and has been studied 
extensively (e.g.~\cite{Alford:1998sd,Lombardo:1999cz,deForcrand:2002hgr,deForcrand:2003vyj,deForcrand:2006pv,deForcrand:2008vr,DElia:2002tig,DElia:2004ani,
Azcoiti:2005tv,Chen:2004tb,Karbstein:2006er,Cea:2006yd,Cea:2007vt,Cea:2010md,Wu:2006su,Nagata:2011yf,Giudice:2004se,DElia:2007bkz,
Cea:2009ba,Alexandru:2013uaa,Cea:2012ev,Conradi:2007be,DElia:2009pdy,Takaishi:2010kc,Cea:2014xva,Cea:2015cya,Bonati:2014kpa,Bonati:2014rfa,
Bonati:2015bha,Bellwied:2015rza,Gunther:2016vcp,DElia:2016jqh,Bornyakov:2017upg,Andreoli:2017zie,Greensite:2014isa,Greensite:2014cxa,
Takahashi:2014rta,Takahashi:2014ofa,Greensite:2017qfl,DElia:2009bzj,deForcrand:2010he,Bonati:2010gi,Philipsen:2014rpa,Wu:2013bfa,
Wu:2014lsa,Nagata:2014fra,Kashiwa:2016vrl,Kashiwa:2015tna,Bonati:2016pwz,Makiyama:2015uwa,Philipsen:2015eya,Cuteri:2015qkq,Bonati:2018fvg,
Kouno:2009bm,Sakai:2009dv,Sasaki:2011wu,Kouno:2011zu,Aarts:2010ky,Rafferty:2011hd,Morita:2011eu,Kashiwa:2011td,Pagura:2011rt,Scheffler:2011te,
Kashiwa:2013rm,Kashiwa:2012xm,Filothodoros:2016txa,Filothodoros:2018pdj}). 
One of the motivations is that it is related by analytic continuation to real chemical potential, and imaginary chemical potential has no sign problem. 
However, the imaginary chemical potential is
also useful for the study of QCD phase transition at zero chemical potential. As mentioned above, it is helpful to study the overall picture of phase diagram when various parameters are changed, such as $\mu_B$.
In particular, the value $\mu_B=\pi$ is special because confinement and deconfinement can be precisely defined at that value. In $\SU(N_c)$ Yang-Mills theory without quarks,
the confinement/deconfinement is characterized by the center symmetry $\BZ_{N_c}$. A well-known problem in QCD with fundamental quarks is that the center symmetry is explicitly broken
and hence confinement/deconfinement is not precisely defined. However, at the special value $\mu_B=\pi$, there is a $\BZ_2$ symmetry~\cite{Roberge:1986mm} 
which can be used as a kind of center symmetry as we will review later in this paper.
We denote this symmetry as $\BZ_2^{\rm center}$ and call the point $\mu_B=\pi$ as the Roberge-Weiss point~\cite{Roberge:1986mm}. Then the phase structure is much more clear at this value because of the
well-definedness of confinement/deconfinement.
See also \cite{Kouno:2012zz,Sakai:2012ika,Kouno:2013zr,Kouno:2013mma,Poppitz:2013zqa,Iritani:2015ara,Kouno:2015sja,Hirakida:2016rqd,Hirakida:2017bye,
Cherman:2017tey,Tanizaki:2017qhf,Tanizaki:2017mtm} for flavor-dependent imaginary chemical potential in which the center symmetry is preserved.

\begin{figure}
\centering
\includegraphics[width=.6\textwidth]{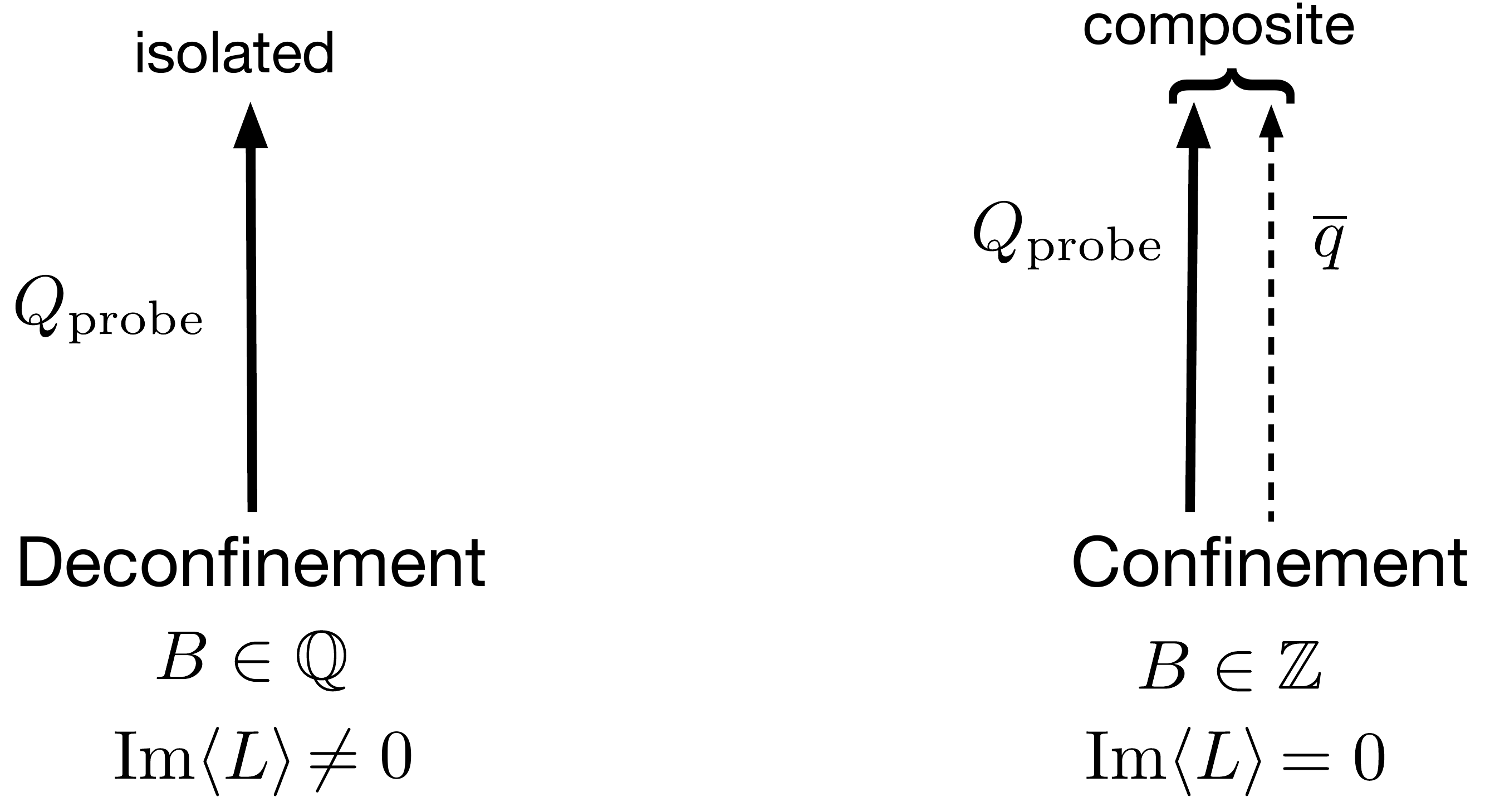}
\caption{ How to define confinement. Here $Q_{\rm probe}$ is the probe quark representing the Polyakov loop, and $\overline{q}$ is a dynamical anti-quark. 
In the confinement phase (Right) the baryon number $B$ is always integer and hence $e^{i \pi B} = \pm 1 \in \BR$, 
while in the deconfinement phase (Left) it is not necessarily integer because quarks have fractional baryon number $1/N_c$ and hence $e^{i \pi B} \in \BC$. \label{fig:confine}}
\end{figure}

Intuitively the reason that confinement is well-defined at $\mu_B=\pi$ is explained as follows, whose details will be discussed later in this paper.
Let $L$ be the Polyakov loop operator (i.e., Wilson loop operator in the direction of the thermal circle $S^1$).
This operator $L$ includes $\mu_B$ (regarded as a background field for baryon symmetry) as well as the usual color gauge fields.
The Polyakov loop may be considered as a world-line of a probe quark (or ``heavy quark" put by hand).
Its vacuum expectation value behaves, intuitively, as
\beq
\vev{L} \sim \exp( - \beta E_q + i \mu_B B)
\eeq
where $\beta=T^{-1}$ is the inverse temperature, $E_q$ is the energy of a probe quark (up to ``bare mass of the heavy quark"), 
$B$ is the baryon number of the probe quark (see below for more discussion), and $\mu_B$
is the imaginary chemical potential. 
In the absence of dynamical quarks, confinement means that an isolated probe quark has an infinite energy $E_q \to +\infty$ and hence confinement (deconfinement) is defined by $\vev{L}=0$ ($\vev{L} \neq 0$).
However, in the presence of dynamical quarks, the probe quark is screened by dynamical anti-quarks as in the right of Figure~\ref{fig:confine} and hence $\vev{L} \neq 0$ in any phase.
However, let us introduce $\mu_B =\pi$. Then the phase of $\vev{L}$ is given by $\exp(i\pi B)$. For gauge invariant composites as in the right of Figure~\ref{fig:confine}
the total baryon number is integer and hence $\exp(i\pi B)$ is real. On the other hand, for an isolated quark as in the left of Figure~\ref{fig:confine} the baryon number is not integer, $B=1/N_c$,
and hence $\exp(i\pi B)$ is imaginary. Therefore, the imaginary part $\Im \vev{L}$ is a good order parameter for a criterion of confinement.
Now, there is a $\BZ_2$ symmetry which changes the direction of the thermal circle $S^1$. Then the Polyakov loop is complex conjugated,
\beq
\BZ_2^{\rm center} : L \to L^*.
\eeq
The $\Im \vev{L}$ is the order parameter of this $\BZ_2^{\rm center}$. We define confinement phase as a phase in which $\BZ_2^{\rm center}$ is unbroken, $\vev{L} =0$.

\begin{figure}
\centering
\includegraphics[width=.8\textwidth]{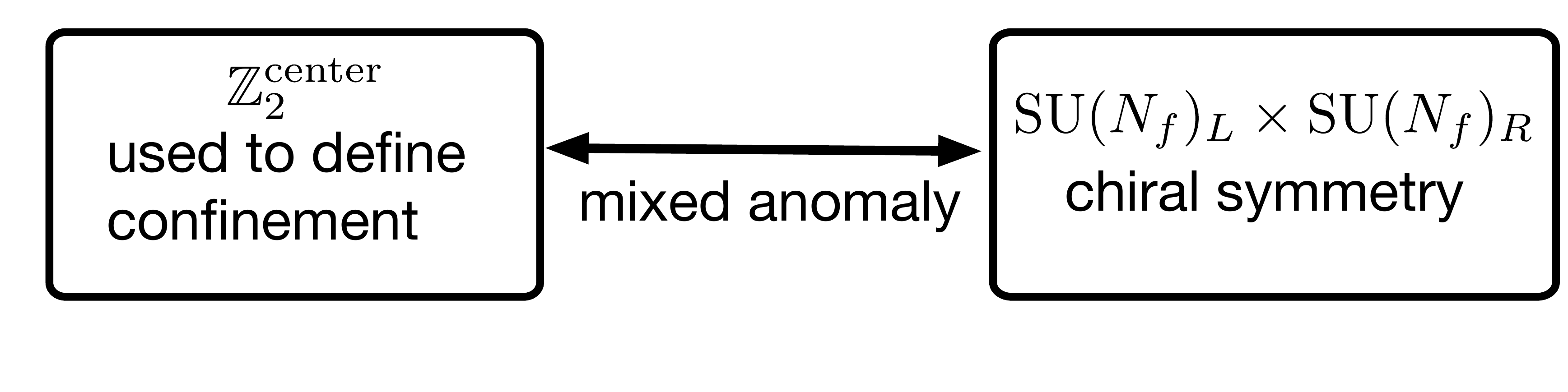}
\caption{ Mixed anomaly between $\BZ_2^{\rm center}$ and $\SU(N_f)_L \times \SU(N_f)_R$ at finite temperature. \label{fig:mixed}}
\end{figure}

We will show in this paper that there is a mixed anomaly between the $\BZ_2^{\rm center}$ symmetry and the chiral symmetry $\SU(N_f)_L \times \SU(N_f)_R$.
This gives a direct strong relation between confinement and chiral symmetry breaking, extending the original results of 't~Hooft.
See Figure~\ref{fig:mixed}.

The existence of the 't~Hooft anomaly puts severe constraints on the nature of phase transition, because the theory at any temperature must have the same 't~Hooft anomaly.
Such an anomaly immediately excludes the simplest scenario of chiral phase transition based on the universality alone, at least at the Roberge-Weiss point $\mu_B=\pi$.
Moreover, we will see that the effect of $\mu_B$ is only sub-leading in the large $N$ expansion. Therefore, the constraints obtained at $\mu_B=\pi$
may have important implications even for the case $\mu_B=0$ if the large $N$ expansion is qualitatively good. Large $N$ analysis is at least qualitatively good in QCD at zero temperature,
so we may hope that it is also useful at finite temperature.

\paragraph{Summary of the results.}
Let us summarize the results of the present paper, which confirm and strengthen the results in \cite{Shimizu:2017asf} by more elementary arguments.

In section~\ref{sec:QCD} we see that finite temperature QCD at the Roberge-Weiss point $\mu_B=\pi$ has a parity anomaly~\cite{Redlich:1983dv,Niemi:1983rq,AlvarezGaume:1984nf,Witten:2016cio}
between $\BZ_2^{\rm center}$ (which will be realized as a parity symmetry in three dimensions after the reduction on the thermal circle $S^1$) and the chiral symmetry $\SU(N_f)_L \times \SU(N_f)_R$.
This is the anomaly shown in Figure~\ref{fig:mixed}.

In section~\ref{sec:chiral} we will reproduce the parity anomaly in the effective theory of pions (i.e. chiral Lagrangian) from the Wess-Zumino-Witten term. In particular, 
a term related to the topological charge of Skyrmions plays the crucial role. The importance of the Skyrmion charge in QCD anomalies was already recognized in \cite{Witten:1983tw,Witten:1983tx}, and
it was used in more sophisticated way in a recent work~\cite{Tanizaki:2018wtg} at zero temperature. 

In section~\ref{sec:transition} we will discuss implications of the anomaly for possible scenarios of QCD phase transition 
at $\mu_B=\pi$, and extrapolate that discussion to $\mu_B=0$ in the large $N$ expansion. At $\mu_B=\pi$, a first order transition may be the most natural scenario of $\SU(N_c)$ QCD phase
transition for generic flavor numbers $N_f \lsim N_c$, although the anomaly itself allows more exotic scenarios such as a deconfined $\U(1)$ gauge field, chiral symmetry breaking in deconfining phase, and so on.
If it is a first order transition, then the result may be unchanged for $\mu_B=0$ as far as the large $N$ analysis is qualitatively valid.

\section{Anomaly of QCD Lagrangian}\label{sec:QCD}

We consider the standard QCD-like theories with general color and flavor numbers $N_c$ and $N_f$.
It is the $\SU(N_c)$ gauge theory with $N_f$ flavors of quark fields $\Psi $ in the fundamental representation of $\SU(N_c)$.
The Lagrangian is
\beq
\CL = -\frac{1}{2g^2} \tr_c ((F_C)_{\mu\nu}(F_C)^{\mu\nu}) + \overline{\Psi}\gamma^\mu D_\mu \Psi,
\eeq
where $(F_C)_{\mu\nu}$ is the field strength of the $\SU(N_c)$ gauge fields $A_C$,\footnote{Gauge fields are taken to be anti-hermitian, e.g. $F_{\mu\nu}^\dagger = - F_{\mu\nu}$ throughout the present paper.}
$\gamma^\mu$ are gamma matrices with $\{\gamma_\mu, \gamma_\nu \} = 2\delta_{\mu \nu}$ in Euclidean space, and $D_\mu=\partial_\mu+(A_C)_\mu$ is the covariant derivative.

The full symmetry group of this theory is a bit complicated, and in this paper we use only partial information.
Including the gauge as well as some of the global symmetry groups, the quark fields are acted by
\beq
H=[\SU(N_c) \times \SU(N_f)_L \times \SU(N_f)_R \times \U(1)_V]/\sD. \label{eq:totsymm}
\eeq
Here $\SU(N_f)_L$ and $\SU(N_f)_R$ are the standard chiral symmetry. 
They act on the left and right handed quarks ${\psi}_L = \frac{(1+\gamma_5)}{2} \Psi$ and $\overline{\psi}_R= \frac{(1-\gamma_5)}{2} \Psi$ as
$\psi_L \mapsto g_L \psi_L$ and $\overline{\psi}_R \mapsto g_R \overline{\psi}_R$
for $g_{L,R} \in \SU(N_f)_{L,R}$. The $\U(1)_V$ acts as $\Psi \to g_V \Psi$ where $g_V \in \U(1)_V$ is a phase factor $|g_V|=1$.
The $\sD$ is a subgroup of the center of the group 
\beq
\sD \subset  \SU(N_c) \times \SU(N_f)_L \times \SU(N_f)_R \times \U(1)_V
\eeq
which acts trivially on the quark fields $\Psi$. More explicitly it is generated by elements $c_1, c_2 \in \sD$ given by
\beq
c_1 &=(e^{2\pi i/N_c}, 1,1, e^{-2\pi i/N_c}), \label{eq:div1}\\
c_2 & =(1, e^{2\pi i/N_f}, e^{2\pi i/N_f}, e^{-2\pi i/N_f}). \label{eq:div2}
\eeq
These $c_1$ and $c_2$ act trivially on the quarks and gluons.

The symmetry group which acts on gauge invariant operators can be obtained by omitting the gauge group $\SU(N_c)$ in $H$.
We get
\beq
G=[ \SU(N_f)_L \times \SU(N_f)_R \times \U(1)_B]/\sC. \label{eq:glob}
\eeq
Here, the baryon number symmetry $\U(1)_B$ is given by
\beq
\U(1)_B = \U(1)_V/\BZ_{N_c}
\eeq
where $\BZ_{N_c}$ is generated by the element $c_1$ above.
The quark fields have charge $1/N_c$ under this $\U(1)_B$.
The $\sC=\BZ_{N_f}$ is generated by $c_2$. In terms of $\U(1)_B$ rather than $\U(1)_V$,
it is given by
\beq
c'_2=(e^{2\pi i/N_f}, e^{2\pi i/N_f}, e^{-2\pi i N_c/N_f}) \in \SU(N_f)_L \times \SU(N_f)_R \times \U(1)_B. \label{eq:div2p}
\eeq
This $c'_2$ acts trivially on all gauge invariant operators.

\subsection{Baryon imaginary chemical potential at the Roberge-Weiss point}\label{sec:RW}
We introduce an imaginary baryon chemical potential $\mu_B$ in the thermal partition function.
Our motivation for introducing it is to make the concept of confinement well-defined as in Figure~\ref{fig:confine}, and
to obtain a concrete 't~Hooft anomaly in the finite temperature situation as in Figure~\ref{fig:mixed}; see Sec.~\ref{sec:intro} for more discussions. 
However, we emphasize that the effect of the imaginary chemical potential is sub-leading in the large $N$ expansion, and hence
our anomaly may also have implications for the case of zero chemical potential. We discuss these points in more detail in Sec.~\ref{sec:transition}.

The thermal partition function in the presence of $\mu_B$ is defined by
\beq
Z(T,\mu_B) = \Tr e^{-\beta H + i \mu_B Q_B} \label{eq:ptn}
\eeq
where $H$ is the Hamiltonian, $T=\beta^{-1}$ is the temperature, $Q_B$ is the baryon charge operator normalized in such a way that quarks have charge $Q_B=1/N_c$,
and the trace is taken over the Hilbert space. Notice that we have normalized $\mu_B$ to be dimensionless which is different from the standard normalization of chemical potentials. 
(The standard chemical potential is given by $T \mu_B$ in our notation.)
Also notice the imaginary unit $i=\sqrt{-1}$ in front of $\mu_B$. Thus this is an imaginary chemical potential.

In Euclidean path integral, the above thermal partition function is obtained by the path integral on $S^1 \times M_3$, where $S^1$
is the thermal circle of circumference $\beta$, and $M_3$ is the spatial manifold (e.g. $M_3=\BR^3$).
In this description, the imaginary chemical potential $\mu_B$ is realized as a background $\U(1)_B$ gauge field $A_B=(A_B)_\mu dx^\mu$.
In this paper we always take Lie algebra generators $T_a$ to be anti-hermitian for mathematical simplicity, and in particular gauge fields $A=T_a A^a$ satisfy
$A^\dagger = - A$. The $A_B$ is pure imaginary in this convention.
Then $\mu_B$ is introduced as
\beq
\mu_B = \int_{S^1} iA_B. \label{eq:icp}
\eeq

Let us notice the following point. The quark fields are coupled to the combination of the dynamical gauge field $A_C$ and the background $\U(1)_B$ field $A_B$ as
$A_C + \frac{1}{N_c} 1_{N_c} A_B$, where $1_{N_c}$ is the unit $N_c \times N_c$ matrix. In particular, the Wilson line around $S^1$ is
\beq
W_{\rm quark}&:={\rm P}\exp\left( - \int_{S^1} (A_C + \frac{1}{N_c} 1_{N_c} A_B) \right) \nonumber\\
&=e^{i\mu_B/N_c}W_{C} \label{eq:Wquark}
\eeq
where
\beq
W_{C}:={\rm P}\exp\left( - \int_{S^1} A_C \right).
\eeq
The $W_{\rm quark}$ is what is relevant for the dynamics of the quarks.

In the absence of the quarks, the pure $\SU(N_c)$ Yang-Mills theory has the so-called center symmetry whose generator acts on $W_C$ as
$W_C \mapsto e^{2\pi i /N_c}W_C$ up to gauge transformations. This is not a symmetry any more in the presence of the quarks, because the quark path integral is not invariant under this transformation.
However, this center symmetry still has some relevance in the presence of the quarks. To see this, let us shift the background $\mu_B$ as
$\mu_B \to \mu_B + 2\pi$. This can be compensated by the shift of the dyamical gauge fields as $W_C \to e^{-2\pi i/N_c} W_C$ so that $W_{\rm quark}$ is invariant.
This means that the thermal partition function is invariant under the shift of the imaginary chemical potential as~\cite{Roberge:1986mm}
\beq
Z(T,\mu_B+2\pi) = Z(T,\mu_B).\label{eq:periodic}
\eeq
The above discussion is related to the fact that in the symmetry group \eqref{eq:totsymm} we divide the group $\SU(N_c) \times \U(1)_V$ by 
the group generated by $c_1$ in \eqref{eq:div1}. The $W_C$ and $e^{i\mu_B/N_c}$ are elements of $\SU(N_c)$ and $\U(1)_V$,
and there is an equivalence relation $(W_C, e^{i\mu_B/N_c}) \sim (e^{2\pi i/N_c} W_C, e^{-2\pi i/N_c} e^{i\mu_B/N_c})$.

Another (equivalent) way of showing the periodicity \eqref{eq:periodic} is as follows.
All gauge invariant states have integer baryon charges, $Q_B \in \BZ$. The $\mu_B$ appears in the thermal partition function as $e^{i \mu_B Q_B}$,
and hence it is invariant under $\mu_B \to \mu_B +2\pi$.

Among the possible values of $\mu_B$, the value $\mu_B = \pi \mod 2\pi$ is special as we explain now, and we call it the Roberge-Weiss point~\cite{Roberge:1986mm}.
We have the time-reversal symmetry (in the Euclidean sense) which changes the $S^1$ coordinate $x^4$ as $x^4 \to -x^4$.
Or more simply, we may combine it with a reflection of one of the coordinates of the space $M_3$ as e.g. $x^3 \to -x^3$.
Then it is a part of the Lorentz group in four dimensions.
We call it a three dimensional reflection symmetry and denote it as $\sR$;
\beq
\sR: (x^3,x^4) \mapsto (-x^3,-x^4).\label{eq:Rx}
\eeq
This changes $\mu_B \to -\mu_B$. However, by the periodicity \eqref{eq:periodic}, the value $\mu_B=\pi$ is invariant under this transformation.
Thus $\sR$ is a symmetry of the theory even in the presence of $\mu_B=\pi$.

The $\sR$ is a symmetry at $\mu_B=0 $ as well as at $\mu_B=\pi$. However, there is something special about $\mu_B=\pi$. 
The symmetry $\sR$ can be used as a criterion of confinement at the Roberge-Weiss point $\mu_B=\pi$ by the following reason.
The $\sR$ is a symmetry at $\mu_B=\pi$ due to the periodicity \eqref{eq:periodic}, and this periodicity is realized by using the shift 
$W_C \to e^{2\pi i/N_c} W_C $. Then, the $\sR$ essentially uses the center symmetry of the gluonic degrees of freedom.
Therefore, it can be used to define confinement. 

Under this symmetry $\sR$, the Wilson line $W_{\rm quark}$ transforms up to gauge transformation as
\beq
\sR: W_{\rm quark} \mapsto W_{\rm quark}^\dagger \label{eq:WunderR}
\eeq
in addition to the coordinate change $x^3 \to -x^3$. This is because the direction of the integration in \eqref{eq:Wquark} is changed by $x^4 \to - x^4$.
Therefore, the imaginary part of the Polyakov loop operator 
\beq
L = \tr_c W_{\rm quark}
\eeq
is an order parameter of the symmetry breaking. 
It is spontaneously broken in deconfinement phase, and preserved in cofinement phase.
Indeed, it is spontaneously broken at high temperature phase~\cite{Roberge:1986mm}
as we will see in Sec.~\ref{sec:transition}. On the other hand, in the low temperature limit $T \to 0$,
the $\sR$ is just a part of the four dimensional Lorentz symmetry as mentioned above and it is preserved.

More intuitively, we may understand the physical situation as follows.
The Polyakov loop $L = \tr_c W_{\rm quark}$ can be regarded as a world-line of a ``heavy quark" which is introduced as a probe.
In deconfinement phase, its vacuum expectation value is given as $\vev{L}  \sim \exp( -\beta E_q + i \mu_B/N_c)$,
where $E_q < +\infty$ is the energy of the color fluxes created by the probe quark.
The appearance of $1/N_c$ is due to the fact that the heavy probe quark is given the baryon number $1/N_c$.
Because of $\exp( i \mu_B/N_c)$, there is a nonzero imaginary part of $\vev{L}$ in the deconfinement phase.
On the other hand, in confinement phase, the probe quark is combined with a dynamical anti-quark to make
a color singlet state. (See Figure~\ref{fig:confine} of Sec.~\ref{sec:intro}.)
This can be interpreted as a meson which consists of the heavy probe quark and a light dynamical anti-quark.
The probe quark can also be combined with $N_c -1$ dynamical quarks to make a color singlet. This is a baryon consisting of the heavy quark and $N_c-1$ light quarks.
Of course there are other possibilities, but the point is that these color singlet states always have integer baryon charges $Q_B \in \BZ$.
Therefore, $\exp(i \mu_B Q_B)$ is just a sign $(-1)^{Q_B}$ at the Roberge-Weiss point $\mu_B=\pi$ and the imaginary part of $\vev{L}$ is zero.
In this way, we can use $\Im \vev{L}$ (or more precisely the symmetry $\sR$) as a criterion of confinement/deconfinement.

\subsection{Parity anomaly in three dimenions}
In this subsection, we will show that in the finite temperature QCD with $\mu_B=\pi$, there is a parity anomaly.
Let us briefly recall the parity anomaly in three dimensions~\cite{Redlich:1983dv,Niemi:1983rq,AlvarezGaume:1984nf,Witten:2016cio}. 
Suppose that we have a fermion $\psi$ coupled to an $\SU(N_f)$ background gauge field $A$ in the fundamental representation with the Lagrangian
\beq
\CL= \overline{\psi} \sigma^i D_i \psi +m \overline{\psi}\psi
\eeq 
where $D_i = \partial_i +A_i$ is the covariant derivative in three dimensions, and $\sigma^i$ are three dimensional gamma matrices (e.g. Pauli matrices).

On the fermion, we can define a reflection $x^3 \to -x^3$ as
\beq
\sR: ~&\psi(x^1,x^2,x^3) \mapsto i\sigma^3 \psi(x^1,x^2, - x^3), \nonumber \\
&  \overline{\psi}(x^1,x^2,x^3) \mapsto  \overline{\psi}(x^1,x^2, - x^3) i\sigma^3. \label{eq:pin-}
\eeq
It is easy to check that the kinetic term is invariant under this $\sR$. On the other hand, the mass parameter changes the sign as $m \to -m$.
Therefore, if $m=0$, the theory has the reflection symmetry $\sR$ at the classical level.

Let us consider it at the quantum level.
Including the Pauli-Villars regulator contribution,
the fermion path integral is given by
\beq
Z_\psi = \frac{\det(\sigma^i D_i + m) }{\det(\sigma^i D_i +M)},
\eeq
where $M \to \infty$ is the regulator mass. Now, we can see that even if $m=0$, the reflection $\sR$ is violated by the regulator mass $M$.
Assuming that the gauge field $A$ is topologically trivial, the change of the partition function before and after the action of $\sR$ is given 
by~\cite{Redlich:1983dv,Niemi:1983rq,AlvarezGaume:1984nf,Witten:2016cio}
\beq
 \frac{\det(\sigma^i D_i -M) }{\det(\sigma^i D_i +M)} \to \exp\left( {\rm CS}(A)  \right). \label{eq:parityanom}
\eeq
Here ${\rm CS}(A)  $ is the Chern-Simons invariant
\beq
{\rm CS}(A)= -\frac{i}{4\pi} \int_{M_3} \tr_f ( A dA + \frac{2}{3}A^3),
\eeq
where the trace is taken in the fundamental representation of $\SU(N_f)$.
This is the famous parity anomaly. More precise treatment~\cite{AlvarezGaume:1984nf,Witten:2016cio} requires the Atiyah-Patodi-Singer $\eta$-invariant~\cite{Atiyah:1975jf,Fukaya:2017tsq},
but we only consider the above rough version in this paper.

If there are $K$ copies of the fermion in the fundamental representation of $\SU(N_f)$, the above anomaly becomes $K {\rm CS}(A)$.
We may try to cancel this anomaly by adding a local counterterm to the action given by
\beq
\frac{K}{2}{\rm CS}(A). 
\eeq 
This term is odd under $\sR$ and changes by $[(-K/2 ) - (K/2)]{\rm CS}(A)=-K{\rm CS}(A)$ which cancels the anomaly \eqref{eq:parityanom}.
This counterterm has no problem if $K$ is even, $K \in 2\BZ$. However, if $K$ is odd, it is
not gauge invariant. Under the gauge transformation
\beq
A \to A^g=g^{-1}Ag +g^{-1} dg
\eeq
the ${\rm CS}$ changes as
\beq
{\rm CS}(A^g)={\rm CS}(A)+ 2\pi i \cdot \frac{1}{24\pi^2} \int_{M_3} \tr_f (g^{-1} dg)^3.
\eeq
The integral $\frac{1}{24\pi^2} \int_{M_3} \tr_f (g^{-1} dg)^3$ is known to be integer (for topologically trivial $\SU(N_f)$ bundle) and it can take the unit value $1 \in \BZ$.
For such $g$, the term $\frac{K}{2}{\rm CS}(A)$ changes by $\pi i K$, so it is not gauge invariant modulo $2\pi i$ if $K$ is odd.
Therefore, we conclude that the true anomaly is characterized by
\beq
K \mod 2
\eeq
which cannot be cancelled by local counterterms.

We want to consider the parity anomaly of the theory which is obtained from the four dimensional QCD after the compactification on the thermal circle $S^1$.
The four dimensional symmetry \eqref{eq:Rx} is reduced to a reflection symmetry $x^3 \to -x^3$ in three dimensions.
Weyl fermions in four dimensions becomes Dirac fermions in three dimensions, and the transformation \eqref{eq:Rx} in four dimensions acts on 
the fermions as in \eqref{eq:pin-} to the massless fields. (See the next paragraph about which fields are massless.)
Massive Kaluza-Klein modes do not contribute to the anomaly, so we can neglect them for the purpose of computing the anomaly.

For simplicity, we compute the anomaly when the configuration of the $\SU(N_c)$ gauge field $A_C$ preserves the symmetry $\sR$; 
such $A_C$ are not at the potential minima (see Sec.~\ref{sec:transition}),
but the anomaly is expected to be independent of $A_C$ because there is no anomaly involving $A_C$ and hence computations for any $A_C$ should give the general result. 

For a configuration of $A_C$ which preserves $\sR$, the Wilson line $W_{\rm quark}$ must be hermitian $W_{\rm quark}^\dagger=W_{\rm quark}$ 
because of \eqref{eq:WunderR}. In addition, it is a unitary matrix because it is defined as a holonomy of the gauge field. 
Then, up to gauge transformations, $W_{\rm quark}$ must be
a diagonal matrix with the eigenvalues either $+1$ or $-1$. Let $K$ be the number of the negative eigenvalue $-1$. Because $W_{\rm quark}=e^{i \mu_B/N_c} W_C$
and $W_C \in \SU(N_c)$, we have 
\beq
(-1)^K=\det(W_{\rm quark})=e^{i\mu_B}
\eeq
and hence $K = \mu_B/\pi \mod 2$.

We are considering the thermal partition function, and hence the fermions have the anti-periodic boundary condition on $S^1$
if the gauge fields are trivial. However, for the components of the fermions coupled to the eigenvalue $(-1)$ of $W_{\rm quark}$,
the gauge Wilson line gives an additional anti-periodicity~\footnote{
Locally, we can take a temporal gauge in which $A_4=0$. Then, globally the effect of the gauge field is represented by additional factor of $W_{\rm quark}$ in the boundary condition.  }
and it cancels against the original anti-periodicity so that these fermions have the periodic boundary condition on $S^1$.
Therefore, there are $K$ massless fermions in three dimensions from each of the left-handed fermion 
$\psi_L = \frac{(1+\gamma_5)}{2} \Psi$ and the right-handed fermion $\overline{\psi}_R= \frac{(1-\gamma_5)}{2} \Psi$.
They are coupled to the background fields $A_L$ and $A_R$ for the chiral symmetries $\SU(N_f)_L$ and $\SU(N_f)_R$, respectively, and
contribute to the parity anomaly in the three dimensional space $M_3$ after the compactification on $S^1$.

If we preserve the symmetry $\sR$ 
by introducing an appropriate counterterm, 
the gauge invariance is broken by an amount characterized by
\beq
\frac{K}{2} ( {\rm CS}(A_L) - {\rm CS}(A_R)),\label{eq:PA}
\eeq
where the coefficients of ${\rm CS}(A_L)$ and ${\rm CS}(A_R)$ are meaningful only modulo 1.

In particular, when $\mu_B=\pi$, the $K$ is odd as discussed above. Therefore, we get a parity anomaly given by \eqref{eq:PA}.
Notice that the chiral symmetry is essential because this anomaly vanishes if we take $A_L = A_R$.

In summary, we found the following mixed anomaly. We define $\BZ^{\rm center}_2$ symmetry\footnote{Strictly speaking,
the $\sR$ generates $\BZ_4$ when it acts on fermions. This symmetry is embedded in $\Pin^-(3)$ group which is obtained from the dimensional reduction of the Lorentz group $\Spin(4)$ in four dimensions.} 
as generated by $\sR$,
\beq
\BZ^{\rm center}_2 = \{ 1, \sR \}. \label{eq:z2center}
\eeq
As discussed above, this symmetry involves the center symmetry of the gluonic degrees of freedom, and hence it can be used to define confinement
even in the presence of dynamical quarks. 
Then there is a mixed anomaly between $\BZ^{\rm center}_2$ and the chiral symmetry $\SU(N_f)_L \times \SU(N_f)_R$
as in Figure~\ref{fig:mixed} of Sec.~\ref{sec:intro}.
Compared with the original 't~Hooft anomaly of the chiral symmetry at zero temperature, our anomaly gives more direct relation
between the two important concepts in QCD: confinement and chiral symmetry breaking.

\section{Anomaly of chiral Lagrangian}\label{sec:chiral}

In this section, we study the effective theory of the Goldstone bosons of chiral symmetry breaking $\SU(N_f)_L \times \SU(N_f)_R \to \SU(N_f)$.
We reproduce the parity anomaly which was found in the previous section.
The Wess-Zumino-Witten (WZW) term plays the crucial role.

Let us recall the effective Lagrangian of Goldstone fields, which we call the chiral Lagrangian.
We represent the Goldstone fields by an $N_f \times N_f$ unitary matrix $U \in \SU(N_f)$.
In the following, we simply call the Goldstone bosons $U$ as pions, although the flavor number $N_f$ is arbitrary in our discussions.
The (Euclidean) effective action of the pion field is given by
\beq
S_{\rm 4d}= \int d^4x \frac{1}{2} f_\pi^2 \tr (D^\mu U^\dagger D_\mu U) + S_{\rm WZW} \label{eq:4dchiral}
\eeq
where $f_\pi$ is the pion decay constant (whose normalization is irrelevant in the present paper), 
$D_\mu$ is the covariant derivative in the presence of background fields for global symmetries,
and $S_{\rm WZW}$ is the Wess-Zumino-Witten (WZW) term which we discuss in detail below.

\subsection{The Wess-Zumino-Witten term}\label{sec:WZW}

Here we would like to describe the WZW term. 
The massless QCD contains the left-handed fermions $\psi_L$ and the right-handed fermions $\overline{\psi}_R$.
Let $\CA_L=(\CA_L)_\mu dx^\mu$ and $\CA_R=(\CA_R)_\mu dx^\mu$ be the gauge fields which are coupled to them. 
Namely, the covariant derivatives on $\psi_L$ and $\overline{\psi}_R$ are given by 
\beq
(D_L)_\mu = (\partial +\CA_L)_\mu , \qquad (D_R)_\mu = (\partial + \CA_R)_\mu, 
\eeq
respectively. They contain background gauge fields for global symmetries as well as dynamical gauge fields for $\SU(N_c)$.
More explicitly, let $A_C$ be the $\SU(N_c)$ dynamical gauge field, $A_L$ and $A_R$ be the $\SU(N_f)_L$ and $\SU(N_f)_R$ background fields,
and $A_B$ be the $\U(1)_B$ background fields. Then, $\CA_L$ and $\CA_R$ are given by
\beq
\CA_{L} &= 1_{N_c} \otimes A_{L} + A_C \otimes 1_{N_f} + \frac{1}{N_c}  1_{N_c} \otimes 1_{ N_f} \cdot A_B \nonumber \\
\CA_R &=  1_{N_c} \otimes A_{R} + A_C \otimes 1_{N_f} + \frac{1}{N_c}  1_{N_c} \otimes 1_{ N_f} \cdot A_B \label{eq:CAdec}
\eeq

The perturbative anomaly is characterized by the anomaly polynomial in six dimensions obtained by the standard descent equations (see e.g. \cite{Weinberg:1996kr}).
The anomaly polynomial 6-form is given as
\beq
I_6 = \frac{1}{3!} \tr \left( \frac{ i \CF_L}{2\pi} \right)^3 -  \frac{1}{3!} \tr \left( \frac{ i \CF_R}{2\pi} \right)^3
\eeq
where $\CF_L = d \CA_L +\CA_L^2$ and $\CF_R = d \CA_R +\CA_R^2$ are the field strength 2-forms, and the trace is taken in the representation of the quarks $\psi_L$ and 
$\overline{\psi}_R$.
Although we have taken $\CA_L$ and $\CA_R$ to include the dynamical $\SU(N_c)$ gauge field,
it disappears from $I_6$ because $\SU(N_c)$ has no anomaly and hence we can neglect the dynamical field of $\SU(N_c)$ in the following discussions.\footnote{
One may think that it is simpler to set the gauge field $A_C$ to be zero in $I_6$ from the beginning. However, the presence of $A_C$ could be important for some purposes.
For example, we can introduce a nontrivial magnetic flux of $A_B$ with unit flux over some cycle.
If the flux is not a multiple of $N_c$,
the introduction of such a flux forces the $\SU(N_c)$ gauge field to be nonzero for mathematical consistency (or more physically by the Dirac quantization condition of fluxes in
each component of the quarks). Therefore,  
we did not set the dynamical field of $\SU(N_c)$ to be zero from the beginning. However, in the present paper we only consider topologically trivial bundles and hence this subtlety does not matter. }

Let $U$ be the unitary matrix of the pion field. We can mathematically describe it by using the matrices $(U_L, U_R) \in \SU(N_f)_L \times \SU(N_f)_R$
and the hidden local symmetry $\SU(N_f)_H$~\cite{Bando:1984ej} as follows. 
The symmetry elements $ g_L \in \SU(N_f)_L$, $g_R \in \SU(N_f)_R$ and $g_H \in \SU(N_f)_H$ act on $(U_L, U_R)$ as
\beq
U_L  \to g_L^{-1} U g_H, \qquad U_R \to   g_R^{-1} U g_H.
\eeq
Here, $g_L$ and $g_R$ are elements of global symmetries, but $g_H$ is a gauge symmetry transformation.
The matrix which is gauge invariant under the hidden local symmetry $\SU(N_f)_H$ is given by
\beq
U=U_L U_R^\dagger. \label{eq:ULR}
\eeq
This is the representation of the pion field $U$ as a field whose target space is the coset space $[\SU(N_f)_L \times \SU(N_f)_R]/\SU(N_f)$ associated to the spontaneous symmetry breaking
$\SU(N_f)_L \times \SU(N_f)_R \to \SU(N_f)$.

We use $(U_L, U_R)$ to transform $(\CA_L, \CA_R)$ as
\beq
\CA_L^{U_L}:&= U_L^{-1} \CA_L U_L + U_L^{-1} d U_L, \\
\CA_R^{U_R}:&= U_R^{-1} \CA_R U_R + U_R^{-1} d U_R.
\eeq
Then, we can see that $\CA_L^{U_L}$ and $\CA_R^{U_R}$ transform in the same way under gauge transformations. 
The fields $\CA_L$ and $\CA_R$ transform as
\beq
\CA_L &\to g_L^{-1} \CA_L g_L + g_L^{-1} d g_L, \\
\CA_R &\to g_R^{-1} \CA_R g_R + g_R^{-1} d g_R.
\eeq
Then by a straightforward computation we see that 
\beq
\CA_{L,R}^{U_{L,R}} \to g_H^{-1} \CA_{L,R}^{U_{L,R}} g_H + g_H^{-1} d g_H,
\eeq
while they are invariant under $\SU(N_f)_L \times \SU(N_f)_R$.
It is also obvious that $\CA_L^{U_L}$ and $\CA_R^{U_R}$ transform in the same way under the color gauge group $\SU(N_c)$ and the baryon number symmetry $\U(1)_B$.
Therefore, $\CA_L^{U_L}$ and $\CA_R^{U_R}$ transform in the same way under any gauge transformation. 
Mathematically, this means that they are connections of the same bundle whose structure group is $\SU(N_c) \times \SU(N_f)_H \times \U(1)_B$ (up to global structure).

Now we can describe the WZW term. From the above fact that $\CA_L^{U_L}$ and $\CA_R^{U_R}$ transform in the same way, it makes mathematical sense
to consider a one-parameter family of gauge fields $\CA_t$ for $t \in [0,1]$ such that $\CA_{t=1} = \CA_L^{U_L} $ and $\CA_{t=0}=\CA_R^{U_R}$. Explicitly, we may just take
\beq
\CA_t = \CA_R^{U_R} + t (\CA_L^{U_L}  - \CA_R^{U_R}) 
\eeq
although it is not necessary.
Now let $\CF_t = d \CA_t +\CA_t^2$ be the field strength of $\CA_t$. By using $\CF_L^{U_L} = U^{-1}_L \CF_L U_L$
and $\CF_R^{U_R} = U^{-1}_R \CF_R U_R$, we can rewrite the 6-form $I_6$ as
\beq
I_6 &= \frac{1}{3!} \tr \left( \frac{ i \CF_L^{U_L}}{2\pi} \right)^3 -  \frac{1}{3!} \tr \left( \frac{ i \CF_R^{U_R} }{2\pi} \right)^3 \nonumber \\
& = \frac{i^3}{3! (2\pi)^3} \int^1_0 dt \frac{\partial}{\partial t}  \tr (\CF_t^3) \nonumber \\
&= \frac{i^3}{2 (2\pi)^3} \int^1_0 dt  \cdot    \tr(\CF_t^2 D_t(\partial_t \CA_t)) \nonumber \\
&= \frac{i^3}{2 (2\pi)^3} \int^1_0 dt  \cdot  d  \tr(\CF_t^2 \partial_t \CA_t) \nonumber \\
&= d I_5 \label{eq:I6I5}
\eeq
where $D_t= d + \CA_t$ in the third line is the covariant exterior derivative and we have used the Bianch identity $D_t F_t=0$.
We have defined the Chern-Simons 5-form as
\beq
I_5 &= \frac{i^3}{2 (2\pi)^3} \int^1_0 dt   \tr (\CF_t^2 \partial_t \CA_t).  \label{eq:I5}
\eeq
The $d$ in the last line of \eqref{eq:I6I5} is the exterior derivative.

Notice that $I_5$ is manifestly gauge invariant. By using it, the WZW term can be given as follows. Let $M_4$ be the 4-dimensional spacetime, and let $N_5$ be a 5-dimensional manifold
whose boundary is $M_4$, i.e. 
\beq
\partial N_5 = M_4. 
\eeq
Then we define the WZW term as
\beq
-S_{\rm WZW} = 2\pi i \int_{N_5} I_5. \label{eq:defWZW}
\eeq
This is gauge invariant as mentioned above. 
The reason that this definition makes sense will be discussed below.

We have defined the WZW term by using the hidden local symmetry. The advantage of using the hidden local symmetry is that it can be generalized
to other gauge theories, such as $\SO(N_c)$ and $\Sp(N_c)$ gauge theories; see Appendix~C of \cite{Tachikawa:2016xvs}.
However, in the present case, we can fix the gauge associated to the hidden local symmetry $\SU(N_f)_H$ as follows.
The $g_H \in \SU(N_f)_H$ acts on $(U_L, U_R)$ as $(U_L, U_R) \to (U_L g_H, U_R g_H)$, so by taking $g_H=U_R^{-1}$ we can fix the gauge as
\beq
(U_L, U_R) \xrightarrow{\text{gauge fixing}} (U, 1)
\eeq
where $U=U_L U_R^{-1}$ as in \eqref{eq:ULR}.
Because $I_5$ is completely gauge invariant, it is allowed to use this gauge fixing.
We will use it in the following. In particular, $\CA_L^{U_L}=\CA_L^{U}$ and $\CA_R^{U_R}=\CA_R$.

Now we would like to check the dependence of the definition \eqref{eq:defWZW} on $N_5$.
Let $N'_5$ be another manifold with the same boundary $\partial N'_5 = M_4$. Then we can glue $N_5$ and $N'_5$ at the common boundary
to get a closed manifold $X_5 = N_5 \sqcup \overline{N'_5}$, where the overline in $\overline{N'_5}$ basically means orientation flip of the manifold
(see \cite{Freed:2016rqq, Yonekura:2018ufj} for more details.)
We have
\beq
2\pi i \left(  \int_{N_5} I_5 - \int_{N'_5} I_5 \right) = 2\pi i \int_{X_5} I_5.
\eeq
Now we argue that it is independent of the pion field $U$ modulo $2\pi i$. 

Let us take another $U'$, and compare the integral $ \int_{X_5} I_5$ for $U$ and $U'$.
Let $\CA'_t$ be one-parameter family of gauge fields for $t \in [0,1]$ such that $\CA'_{t=0}=\CA_L^U$ and $\CA'_{t=1}=\CA_L^{U'}$.
From the definitions, one can see that the difference of $ \int_{X_5} I_5$ between $U$ and $U'$ is given by
\beq
\frac{i^3}{2 (2\pi)^3} \int_{X_5}  \int_0^1 dt  \tr ((\CF'_t)^2 \partial_t \CA'_t).\label{eq:UUp}
\eeq
where $\CF'_t = d \CA'_t + (\CA'_t)^2$. 

We introduce a six dimensional manifold $Z_6:=[0,1] \times X_5$, where $[0,1]$ is parametrized by $t$.
The exterior derivative on it is given by $\widehat{d} = d + dt \cdot \partial_t$, where $d$ is the exterior derivative on $X_5$.
We can regard $\CA'_t$ as a gauge field on $Z_6$ where $t $ is the coordinate of $[0,1]$.
The field strength of $\CA'_t$ on $Z_6$ is given by, 
\beq
\widehat{\CF}' = \widehat{d} \CA'_t + (\CA'_t)^2 = \CF'_t+ dt \cdot \partial_t \CA'_t,
\eeq
where $\CF'_t$ is the field strength of $\CA'_t$ when it is regarded as a gauge field on $X_5$ for fixed $t$.
We can now rewrite the equation \eqref{eq:UUp} as 
\beq
\frac{i^3}{3! (2\pi)^3} \int_{Z_6}   \tr ( \widehat{\CF}')^3. \label{eq:ch3}
\eeq
Moreover, the gauge field $\CA'_t$ at $t=0$ and $t=1$ only differs by a gauge transformation $U^{-1}U'$ and hence we can glue $t=0$ and $t=1$ to make $Z_6$ a closed manifold.
Then the above integral \eqref{eq:ch3} is integer by Atiyah-Singer index theorem in 6-dimensions.
This proves that the difference of $2\pi i \int_{X_5} I_5$ between $U$ and $U'$ is only integer multiples of $2\pi i$ which is irrelevant in the action.
Therefore, we have shown that $2\pi i  \int_{X_5} I_5 \mod 2\pi i$ is independent of $U$.

By the above discussion, we have shown that our definition of the WZW term $2\pi i \int_{N_5} I_5$ satisfies the condition that
its $U$ dependence does not depend on how to take the 5-dimensional manifold $N_5$ with $\partial N_5 = M_4$.
If it were completely independent of $N_5$, we can say that the WZW term would depend only on $M_4$.
However, the $2\pi i \int_{N_5} I_5$ still depends on how to take the manifold $N_5$ via the fields $\CA_L$ and $\CA_R$.
This is exactly the modern understanding of anomalies in general (see \cite{Witten:2015aba} and references therein). 
Namely, the definition of the path integral depends on manifolds $N_5$,
but the dependence is only through the background fields and this dependence is the anomaly of the corresponding symmetries.
 In other words, the anomaly is characterized by the symmetry protected topological (SPT) phase in five dimensions
whose partition function is given by $\exp(2\pi i \int_{X_5} I_5)$. Therefore, our definition of the WZW term is in accord with the modern understanding of anomalies,
and the WZW term reproduces the perturbative anomaly represented by the anomaly polynomial $I_6$.

Our definition of $2\pi i \int_{N_5} I_5$ is gauge invariant but depends on $N_5$, in accord with the modern understanding. 
However, it is also possible to make it independent of $N_5$ (for topologically trivial cases)
at the cost of introducing gauge non-invariance. That is more close to the old understanding of anomalies. 
See Appendix~A of \cite{Yonekura:2016wuc} for the details of these points.
But we remark that the dependence on $N_5$ is more fundamental when we consider global anomalies.

Finally, let us check that the above definition of the WZW term coincides with the textbook definition when the background fields $\CA_{L,R}$ are zero.
If $\CA_L=\CA_R=0$ (and hence $\CA_L^U=U^{-1}dU$), then by a straightforward computation one can check that \eqref{eq:I5} gives
\beq
2\pi i \int_{N_5} I_5 = \frac{1}{240\pi^2} \int_{N_5} \tr (U^{-1} dU)^5 = \frac{N_c}{240\pi^2} \int_{N_5} \tr_f (U^{-1} dU)^5,
\eeq
where $\tr$ is the trace over both the color and flavor (in the representation of the quarks), and $\tr_f$ is the trace only over the favor. 
This is the standard expression for the WZW term in the absence of the background fields.

\subsection{Baryon imaginary chemical potential in the chiral Lagrangian}

The gauge fields $\CA_L$ and $\CA_R$ are decomposed as \eqref{eq:CAdec}.
By expanding the anomaly polynomial $I_6$, we get a term
\beq
I_6 \supset \frac{i^3}{2(2\pi)^3} F_B \left( \tr_f F_L^2 - \tr_f F_R^2 \right).
\eeq
By the procedure explained above, we see that the 5-form $I_5$ contains a term
\beq
I_5 &\supset \frac{i^3}{(2\pi)^3} \cdot F_B \int_0^1 dt \tr_f \left( F_t \partial_t A_t       \right) \nonumber \\
&:= \frac{i F_B}{2\pi } I_3
\eeq
where $A_t = A_R +t (A_L^U-A_R)$, $F_t = d A_t +A_t^2$ and
\beq
I_3 := \frac{i^2}{(2\pi)^2} \cdot   \int_0^1 dt \tr_f \left( F_t \partial_t A_t       \right) 
\eeq
which satisfies
\beq
d I_3 = \frac{i^2}{2(2\pi)^2}  \left( \tr_f F_L^2 - \tr_f F_R^2 \right).
\eeq

Now we consider the following more specific situation. When we are interested in thermodynamics, we consider
a 4-manifold $M_4$ of the form 
\beq
M_4=S^1 \times M_3, 
\eeq
where $S^1$ is the thermal circle of circumference $\beta=T^{-1}$,
and $M_3$ is a 3-manifold. Moreover, we take the baryon background field with a constant holonomy around $S^1$ as in \eqref{eq:icp},
\beq
\int_{S^1} i A_B = \mu_B.
\eeq
We take all the other fields $A_{L,R}$ and $U$ to depend only on $M_3$ and they are constant in the direction $S^1$.
In this situation, we can take $N_5 = D^2 \times M_3$, where $D^2$ is a two dimensional disk with the boundary $\partial D^2 = S^1$,
and $F_B=dA_B$ has the flux $\int_{D^2} i F_B = \int_{S^1} i A_B = \mu_B$. Therefore, the WZW term is reduced as
\beq
2\pi i \int_{N_5} I_5 \to  i \mu_B \int_{M_3} I_3.  \label{eq:theta}
\eeq

In the case of zero background fields $A_L=A_R=0$, we get $I_3 =  \frac{1}{24\pi^2} \tr_f (U^{-1} dU)^3 $ and
\beq
i \mu_B \int_{M_3} I_3 = i \mu_B \int_{M_3}  \frac{1}{24\pi^2} \tr_f (U^{-1} dU)^3. \label{eq:theta2}
\eeq
The integral $\int_{M_3}  \frac{1}{24\pi^2} \tr_f (U^{-1} dU)^3$ is quantized to be integers in the absence of the background fields.
Thus we get the following conclusion. The imaginary baryon chemical potential $\mu_B$ plays the role of a $\theta$ angle
in the effective theory of the 3-dimensional sigma model described by $U$ which is obtained from the dimensional reduction of the 4-dimensional sigma model.
The quantization implies that $\mu_B$ dependence has a periodicity $\mu_B \sim \mu_B+2\pi$. 

The expression \eqref{eq:theta2} itself could be more easily derived if we use the following fact. In the chiral Lagrangian, the baryon current $J_B$ (which we take to be a 3-form) 
is given by the topological current given by~\cite{Witten:1983tw,Witten:1983tx}~\footnote{The current \eqref{eq:BC} was originally derived from the considerations of the WZW term and the anomaly
as discussed above. So the following discussion is not independent from the discussions given above.}
\beq
J_B =  \frac{1}{24\pi^2} \tr_f (U^{-1} dU)^3 \label{eq:BC}
\eeq
and the baryon charge is given by $Q_B =\int_{M_3} J_B$.
Therefore, the imaginary chemical potential appears as (see \eqref{eq:ptn})
\beq
\exp( i \mu_B Q_B ) = \exp\left(  i \mu_B \int_{M_3}  \frac{1}{24\pi^2} \tr_f (U^{-1} dU)^3 \right).
\eeq
However, in the above discussions we have derived the more complete expression \eqref{eq:theta} which incorporates nonzero background fields.
This is essential for our purposes below.

\subsection{Three dimensional effective field theory and parity anomaly}
Let us summarize what we have found above.
In four dimensions, we have the chiral Lagrangian given by \eqref{eq:4dchiral}.
We put the system at finite temperature $T = \beta^{-1}$ with the imaginary baryon chemical potential $\mu_B$.
This means that we compactify the theory on $S^1$ with circumference $\beta$ and the holonomy $\int_{S^1} i A_B = \mu_B$.
If the temperature is lower than any critical temperature, we obtain the three dimensional effective theory given by
\beq
-S_{\rm 3d} = -\int_{M_3} d^3x \frac{1}{2} f_\pi^2 \beta \tr (D^\mu U^\dagger D_\mu U)  +   i \mu_B \int_{M_3} I_3.
\eeq
The second term plays the role of a $\theta$ angle $\theta = \mu_B$ for the three dimensional sigma model described by $U \in \SU(N_f)$.

The term $\int_{M_3}I_3$ is completely gauge invariant.
It has the following alternative description which will be convenient for later purposes. 
Let $A_t$ be one-parameter family of gauge fields on $M_3$ with $A_{t=0}=A_R$ and $A_{t=1}=A_L^U$.
We can consider it as a gauge field on a 4-dimensional manifold $L_4 = [0,1] \times M_3$ where $[0,1]$ is parametrized by $t$.
The exterior derivative on $L_4$ is denoted as $\widehat{d}=d + dt \cdot \partial_t$, where $d$ is the exterior derivative on $M_3$.
The field strength on $L_4$ is given as
\beq
\widehat{F} = \widehat{d} A_t +A_t^2= d A_t+A_t^2 + dt \cdot \partial_t A_t = F_t + dt \cdot \partial_t A_t,
\eeq
where $F_t$ is the field strength when we regard $A_t$ as a gauge field on $M_3$ for fixed $t$.
Then we have
\beq
\int_{M_3}I_3 &=  \frac{i^2}{(2\pi)^2}  \int_{M_3} \int_0^1 dt \tr_f (F_t  \partial_t A_t) \nonumber \\
&=\frac{i^2}{2(2\pi)^2}  \int_{L_4}  \tr_f (\widehat{F})^2. \label{eq:I3invform}
\eeq

Now we focus our attention to the Roberge-Weiss point
\beq
\mu_B=\pi.
\eeq
At this point, there is a parity (or more precisely reflection or time-reversal) symmetry if the background fields are turned off.
In three dimensions, we consider a reflection $\sR$ of one of the coordinates, say $x^3$, as $x^3 \to -x^3$.
In the full four dimensional manifold, it is actually a rotation of $(x^3, x^4)$ plane where $x^4$ is the direction of the thermal circle $S^1$, as in \eqref{eq:Rx}.
Let us see how this reflection invariance is realized in the chiral Lagrangian. The reflection changes $\mu_B $ as
\beq
\mu_B \to - \mu_B.
\eeq
Therefore, at $\mu_B=\pi$, the change of the chiral Lagrangian is given as
\beq
(\pi i I_3) - (-\pi i I_3) = 2\pi i \int_{M_3} \frac{1}{24\pi^2} \tr_f (U^{-1} dU)^3
\eeq
where in the equality we have set the background fields $A_L$ and $A_R$ to be zero for the moment. (We will soon recover them.) 
Because of the quantization of the topological charge $ \int_{M_3} \frac{1}{24\pi^2} \tr_f (U^{-1} dU)^3 \in \BZ$, the above change 
of the chiral Lagrangian is just integer multiple of $2\pi i$ which does not affect the exponential of the action. 
Therefore, this is a symmetry of the three dimensional effective theory.

Next let us introduce the background fields $A_L$ and $A_R$. The change of the effective action under the reflection $x^3 \to - x^3$ (and $x^4 \to -x^4$)
is given by
\beq
2\pi i I_3 = 2\pi i \cdot \frac{i^2}{2(2\pi)^2}  \int_{L_4}  \tr_f (\widehat{F})^2,
\eeq
where we used \eqref{eq:I3invform}.

First, we note that $2\pi i I_3 $ is independent of $U$ modulo $2\pi i \BZ$. The proof is completely analogous to the proof that $2\pi i \int_{X_5} I_5 \mod 2\pi i $
is independent of $U$ as was shown in Sec.~\ref{sec:WZW}, so we do not repeat it.

Now, to make the expression simpler, we assume that $A_L$ and $A_R$ are topologically trivial. 
We define
\beq
{\rm CS}(A) = -\frac{i}{4\pi} \int_{M_3} \tr_f \left( AdA +\frac{2}{3} A^3 \right).
\eeq
For topologically trivial $A_L$ and $A_R$, the Stokes theorem gives
\beq
2\pi i I_3 &= 2\pi i \cdot \frac{i^2}{2(2\pi)^2}  \int_{L_4}  \tr_f (\widehat{F})^2 \nonumber\\
&= {\rm CS}(A_L^U) - {\rm CS}(A_R) \nonumber \\
& = {\rm CS}(A_L) - {\rm CS}(A_R) \mod 2\pi i \BZ, \label{eq:PformU}
\eeq
where we have used $ {\rm CS}(A_L^U)  =  {\rm CS}(A_L) \mod 2\pi i$.
It is indeed independent of $U$. Moreover, this is nonzero and gives
the anomaly of $\sR$. This anomaly in the sigma model at $\mu_B=\pi$ is somewhat analogous to the anomaly found
in gauge theories with topological $\theta$ angle at $\theta=\pi$~\cite{Gaiotto:2017yup,Komargodski:2017dmc}, because $\mu_B$ plays the role of a $\theta$ angle in our sigma model.

The above computation \eqref{eq:PformU} reproduces the parity anomaly \eqref{eq:parityanom}. As in Sec.~\ref{sec:QCD},
we can also introduce a counterterm \eqref{eq:PA} with $K=1$ to recover the invariance under $\sR$, but then the gauge invariance is spoiled.
We have confirmed that QCD at very high and very low temperatures indeed satisfy the 't~Hooft anomaly matching condition.
This anomaly must be matched also at intermediate temperatures, and that gives constraints on the nature of QCD phase transition.

\section{Implications for QCD phase transition}\label{sec:transition}
In the previous sections, we found an 't~Hooft anomaly of global symmetries. It is a mixed anomaly between $\BZ^{\rm center}_2=\{1,\sR\}$ defined in \eqref{eq:Rx} and $\SU(N_f)_L \times \SU(N_f)_R$.
Here we would like to discuss implications of the anomaly for the QCD phase transition.

\subsection{The effect of imaginary chemical potential}
To make the concept of confinement well-defined, we have introduced the imaginary chemical potential $\mu_B=\pi$.
What we are most interested in is the zero chemical potential case $\mu_B=0$,\footnote{It is also extremely interesting to consider real chemical potential. 
For a study of finite density QCD by using 't~Hooft anomaly in a different set up, see \cite{Tanizaki:2017mtm,Tanizaki:2018wtg}.} although $\mu_B=\pi$ is theoretically very interesting because of the well-definedness of confinement. 
Therefore we want to estimate the effect of nonzero $\mu_B$.

In the large $N_c$ expansion, the effect of $\mu_B$ is suppressed by powers of $1/N_c$.
Let us define the thermal free energy density by
\beq
F(T,\mu_B)=- \frac{1}{\beta V_3}\log Z(T,\mu_B),
\eeq
where $V_3$ is the volume of the spatial manifold $M_3$.
First of all, gluons contribute to the thermal free energy $F$ by the order of $N^2_c$ 
and quarks contribute $N_c N_f$. They are just the numbers of degrees of freedom of these fields.
The $\mu_B$ is coupled to quarks as
\beq
 \CL \supset \frac{\mu_B}{N_c} \bar{\Psi}\gamma^4 \Psi.
\eeq
Here we used the fact that the baryon number of the quarks is $1/N_c$. Moreover, the reflection along the $S^1$ direction $x^4 \to -x^4$
implies that the free energy depends only on even powers of $\mu_B$ (assuming it has power expansion around $\mu_B=0$), because $F(T,\mu_B)$ must be invariant under $\mu_B \to -\mu_B$.
Combining these facts with the standard large $N$ counting, we can estimate that the effect of $\mu_B$ to the free energy is of order
\beq
N_c N_f \left( \frac{\mu_B}{N_c} \right)^2.
\eeq
This estimate is valid even if $N_f$ is comparable to $N_c$.
Therefore, the free energy has the following $N_c$ and $N_f$ dependence:
\beq
F \sim   N_c^2+N_cN_f + \frac{N_f}{N_c} \mu_B^2, \label{eq:count1}
\eeq
where the first term comes from gluons, the second term comes from quarks, and the third term is the effect of $\mu_B$.
This is the situation at high enough temperatures, and we will give more explicit values later.
We can see that the term containing $\mu_B^2$ is suppressed by multiple powers of $1/N_c$.

At low enough temperatures after confinement, the only light degrees of freedom are the pions (i.e. Goldstone bosons of the chiral symmetry breaking).
Any particle which has nonzero baryon charge has a mass larger than the lowest baryon mass (i.e. the proton mass in the real world) which we denote as $M_B$.
Then the free energy is of the form
\beq
F \sim N_f^2 + N_{B}  e^{- M_B/T}(e^{i \mu_B} +\text{c.c.})
\eeq
where $N_{B}$ is the number of degrees of freedom of the lowest mass baryon.
The baryon mass behaves as $M_B \sim N_c \Lambda$ in the large $N_c$ expansion~\cite{Witten:1979kh}, where $\Lambda$ is the typical scale of the strong dynamics.
As far as the temperature is below this typial scale $T \lsim \Lambda$, the Boltzmann suppression factor $e^{- M_B/T}$ gives an exponential suppression.
Therefore, the term containing $\mu_B$ is exponentially suppressed in the large $N_c$ limit.

We conclude that at any temperature, the effect of $\mu_B$ is suppressed at least in the large $N_c$ limit.
In QCD with $N_c=3$ and $2 \lsim N_f \lsim 3$, the large $N_c$ analysis works at least qualitatively at zero temperature, explaining many phenomena
in QCD phenomenology (see e.g. the first few sections of \cite{Witten:1979kh} for a review). 
So it is reasonable to hope that $\mu_B$ does not
have significant effects on the thermodynamics in the case of $N_c \gsim 3$ and $N_f \lsim N_c$.

Let us give a bit more numerical estimate in the high temperature limit. 
As a preparation, we define a function $f(T,\mu)$ as
\beq
f(T,\mu) &= \pi^2 T^4\left( - \frac{7}{8} \cdot\frac{1}{45} + \frac{1}{12}\left( \frac{\mu}{\pi} \right)^2 - \frac{1}{24} \left( \frac{\mu}{\pi} \right)^4 \right) ~~~ \text{for}~ -\pi \leq \mu \leq \pi, \\
f(T, \mu)& = f(T,\mu+2\pi) .
\eeq
This function is not completely smooth at $\mu=\pi \mod 2\pi$.

Suppose that we have a massless particle and its anti-particle which are coupled to Wilson line with $\int_{S^1} i A = \mu$. Then the free energy
of that particle is, neglecting interactions, given by
\beq
\left\{ \begin{array}{cc}
\text{boson : } & -f(T, \mu+\pi) \\
\text{fermion : } & f(T,\mu)
\end{array}
\right.\label{eq:FE}
\eeq
When $\mu=0$, one can check that it reproduces the standard free energy. If the anti-particle is the same as the particle itself, we divide it by 2 and set $\mu=0$.
The difference between $\mu+\pi$ for bosons and $\mu$ for fermions comes from the fact that fermions have anti-periodic boundary conditions.
The periodic boundary condition for fermions can be realized by replacing $\mu$ by $\mu+\pi$, although 
for our applications we only need the anti-periodic case.\footnote{As a check, for the periodic case,
the contribution of a boson is exactly negative of that of a fermion, which should be the case if the theory is supersymmetric.}

Let us consider the case of QCD. Suppose that the gauge field $A_C + \frac{1}{N_c} 1_{N_c} A_B$ which appears in \eqref{eq:Wquark}
is diagonal and is given by
\beq
\int_{S^1} i(A_C + \frac{1}{N_c} 1_{N_c} A_B) = \diag (a_1, \cdots, a_{N_c}).
\eeq
The $a_i$ must satisfy the constraint
\beq
\sum_{i=1}^{N_c} a_i = \mu_B + 2\pi \BZ \label{eq:sumuB}
\eeq
because ${\rm P}\exp ( - \int_{S^1} A_C) \in \SU(N_c)$. Then the free energy of QCD for a fixed set $\{ a_i \}$ is given by
\beq
F(T, \{ a_i\})=  F_{\rm gluon}(T, \{ a_i\}) + F_{\rm quark}(T, \{ a_i\})
\eeq
where the gluon contribution $F_{\rm gluon}(T, \{ a_i\}) $ and the quark contribution $F_{\rm quark}(T, \{ a_i\})$ are given by (see \eqref{eq:FE})
\beq
F_{\rm gluon}(T, \{ a_i\}) &= -(N_c-1)f(T, \pi) -\sum_{ i \neq j} f(T, a_i - a_j+\pi) \\
F_{\rm quark}(T, \{ a_i\}) &= 2N_f \sum_i f(T,a_i).
\eeq
The actual free energy (at the leading order of perturbation theory) is given by minimizing this free energy under the condition \eqref{eq:sumuB}.

The gluon part has the minimum at
\beq
a_1=\cdots = a_{N_c} = \frac{\mu_B + 2\pi n}{N_c}
\eeq
where $n \in \BZ$ is an integer. Notice that the center symmetry acts as $n \to n+1$ and hence different values of $n$ are related by the center symmetry in the absence of the quarks.

At least in the large $N_c$ limit we just need to minimize the quark contribution with respect to $n$.
If $0 \leq \mu_B <\pi$ we find that there is a unique minimum at $n=0$. However, if $\mu_B=\pi$, there are two minima at $n=-1$ and $n=0$.
These two minima are related by the symmetry $\sR$ given in \eqref{eq:Rx}. Therefore, the symmetry $\BZ^{\rm center}_2=\{1,\sR \}$ is spontaneously broken at high temperatures.

The free energy at $n=0$ is given by
\beq
F(T, \mu_B)= F_{\rm gluon}(T) + F_{\rm quark}(T, \mu_B)
\eeq
where 
\beq
F_{\rm gluon}(T)  &= -(N^2_c-1)f(T, \pi)  = -(N_c^2-1) \frac{\pi^2}{45}T^4       
\eeq
and
\beq
F_{\rm quark}(T,\mu_B ) &= 2N_fN_c  f(T, \frac{\mu_B}{N_c}) \nonumber \\
&=  -2N_fN_c \frac{\pi^2}{45} T^4\left( \frac{7}{8} - \frac{15 \mu_B^2}{4 \pi^2 N_c^2} + \frac{15 \mu_B^4}{8 \pi^4 N_c^4}  \right).
\eeq
These are really of the form \eqref{eq:count1} which we argued based on the large $N_c$ counting.

Taking the ratio of the free energy at $\mu_B=0$ and $\mu_B=\pi$, we get
\beq
r:=1-\frac{F(T, \mu_B=\pi)}{F(T, \mu_B=0)} =  \frac{15 N_f ( 1 - 2^{-1} N_c^{-2}   ) }{ 2N_c[(N_c^2-1) + \frac{7}{4} N_f N_c ] }.
\eeq
This $r$ may be regarded as an estimate of the effect of the baryon imaginary chemical potential at $\mu_B=\pi$.
As an example, if we set $(N_c, N_f)=(3,2)$, we get $r \simeq 0.25$. 
For very large $N_c \gg N_f$ we have $r \simeq \frac{15}{2} \frac{N_f}{N_c^3}$.

On the other hand, for $(N_c,N_f)=(2,2)$, we get $r \simeq 0.65$. In this case, a more serious problem is as follows. 
For $\SU(2)$ gauge group, the fundamental and anti-fundamental representations are the same and the chiral symmetry is enhanced to $\SU(2N_f)$.
At low temperatures the lightest baryons are actually Goldstone bosons
of spontaneously broken symmetry $\SU(2N_f) \to \Sp(N_f)$. Thus they are massless and $\mu_B$ should have large impact on the dynamics if $N_c=2$.
See also \cite{Hanada:2018zxn} for another reason that $N_c=2$ may be quite different from $N_c \geq 3$.
For $N_c \gsim 3$, we can hope that $\mu_B$ has only sub-leading effects.

\subsection{Possible senarios of QCD phase transition}
Because of the difficulty of analytical studies of QCD phase transition, it is often assumed that the idea of universality can be applied to it based on the chiral symmetry~\cite{Pisarski:1983ms}.
The idea is as follows. We define meson scalar field as
\beq
\Phi = \overline{\Psi} \left( \frac{1+\gamma_5}{2} \right) \Psi = \psi_L \psi_R
\eeq
where $\psi_L$ is the left-handed quarks, $\psi_R$ is (the complex conjugate of) the right-handed quarks, and the color and spin indices are contracted while the flavor 
indices are not. Thus $\Phi$ is an $N_f \times N_f$ matrix on which the chiral symmetry transformation $(g_L, g_R) \in \SU(N_f)_L \times \SU(N_f)_R$ acts as
\beq
\Phi \to g_L \Phi g_R^\dagger.
\eeq
After the chiral symmetry breaking $\vev{\Phi} \neq 0$, this meson field $\Phi$ is represented by the Goldstone field $U$ as $\Phi \propto U$.

Assuming that $\Phi$ is the most important order parameter of the chiral symmetry breaking, we may try to describe the phase transition by 
an effective field theory based on a linear sigma model of $\Phi$,
\beq
\CL_\Phi \sim  \tr_f (\partial_\mu \Phi^\dagger \partial^\mu \Phi) + (T-T_{\rm chiral}) \tr_f (\Phi^\dagger \Phi) + \cdots \label{eq:univ}
\eeq
where ellipses denote terms of higher powers of $\Phi$. Below the critical temperature of chiral phase transition $T_{\rm chiral}$ the mass term is negative, and
$\Phi$ gets a vacuum expectation value (VEV) which spontaneously breaks the chiral symmetry. Above the critical temperature $T \geq T_{\rm chiral}$,
the mass term becomes non-negative and the chiral symmetry is restored. This is the argument based on universality.
If this argument is really true, it is belived that the phase transition for the two flavor case $N_f=2$ is second order phase transition.\footnote{
If the anomalous axial $\U(1)_A$ symmetry is effectively restored, this conclusion may change~\cite{Aoki:2012yj,Fukaya:2017wfq,Chiu:2013wwa,Nakayama:2014sba}.}
If so, small quark masses can make the second order transition to just a cross-over without a clear phase transition. 
This argument is incoorporated in the standard Columbia plot of QCD phase diagram; see e.g. \cite{Fukushima:2010bq} for a review.

Based on the anomaly which we found in the previous sections, we would like to argue that the above picture of QCD phase transition is quite disfavored, if large $N$ expansion is qualitatively good.
Roughly, the argument is that the effective theory \eqref{eq:univ} does not have the parity anomaly and hence fails the anomaly matching condition.
First we consider the case $\mu_B=\pi$ and then discuss the case $\mu_B=0$ later.
We always assume that $N_c \geq 3$ in the following.

\subsubsection{$\mu_B=\pi$}
In the case $\mu_B=\pi$, it is crucial to notice that there is the important symmetry other than the chiral symmetry; the symmetry $\BZ_2^{\rm center}=\{1,\sR\}$ where $\sR$ was defined in \eqref{eq:Rx}.
As explained in Sec.~\ref{sec:RW}, this symmetry $\BZ_2^{\rm center}$ involves 
the center symmetry transformation $W_C \to e^{2\pi i /N_c} W_C$ of the gluonic degrees of freedom, and hence it can be used as a criterion of 
confinement and deconfinement. Namely, if $\BZ_2^{\rm center}$ is spontaneously broken, we consider the phase as a deconfining phase.
Indeed, as seen in the previous subsection, this symmetry is spontaneously broken at very high temperatures.
Near zero temperatures, it is just a part of the four dimensional Lorentz group and hence it is unbroken.
So there must be a phase transition which is a confinement/deconfinement phase transition.

We assume that the chiral symmetry is spontaneously broken at low temperatures. This is a reasonable assumption if the flavor number is not so large, $N_f \lsim N_c$.
Then there is also a chiral phase transition related to the chiral symmetry breaking/restoration. 

We denote the critical temperatures associated to the symmetries $\BZ_2^{\rm center}$ and $\SU(N_f)_L \times \SU(N_f)_R$ as $T_{\rm center}$ and $T_{\rm chiral}$,
respectively. They can coincide as $T_{\rm center} = T_{\rm chiral}$ (and we will argue that this may be the most natural case with a first order transition at that temperature).
However, in principle these critical temperatures can be different. Logically it is also possible to have more than two critical temperatures which are not required by the symmetries,
but we regard such possibilities as unlikely and just consider $T_{\rm center} $ and $ T_{\rm chiral}$.

The situation is summarized in Table~\ref{table:1}.
\begin{table}
\caption{The temperature dependence of symmetry breaking/restoration. Here the ``intermediate $T$" means that the temperature is within the range
$\min(T_{\rm chiral}, T_{\rm center}) \leq T \leq \max(T_{\rm chiral}, T_{\rm center})$. }
\begin{center}
\begin{tabular}{|c|c|c|c|}
\hline
& high $T$ & intermediate $T$& low $T$ \\ \hline
$\BZ_2^{\rm center}=\{1,\sR\}$ & broken &?& unbroken \\ \hline
$\SU(N_f)_L \times \SU(N_f)_R$ & unbroken & ? & broken \\ \hline
\end{tabular}\label{table:1}
\end{center}
\end{table} 
The parity anomaly found in the previous sections is a mixed anomaly 
between $\BZ_2^{\rm center}$ and $\SU(N_f)_L \times \SU(N_f)_R$.

\begin{figure}
\centering
\includegraphics[width=0.7\textwidth]{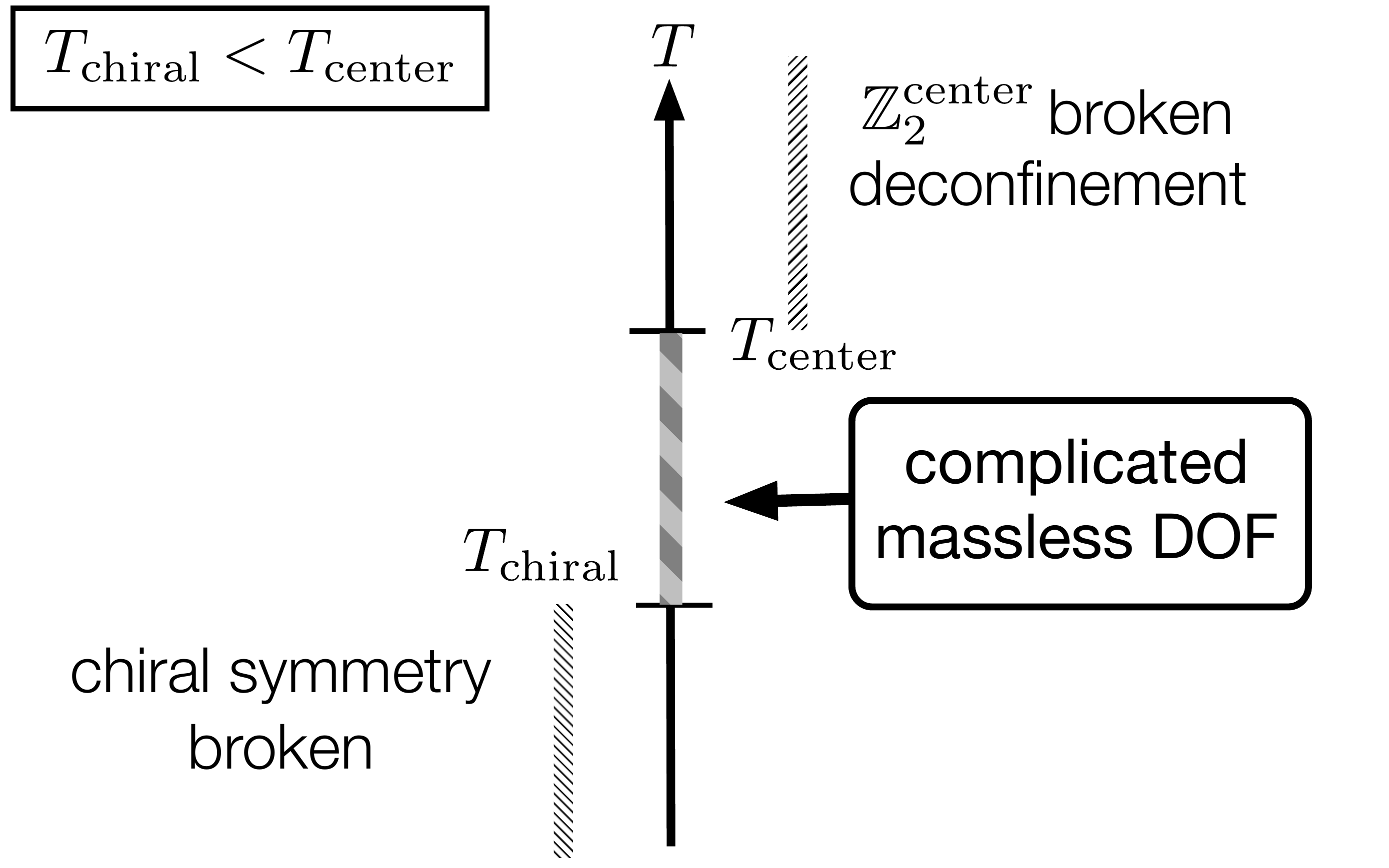}
\caption{The case  $T_{\rm chiral} < T_{\rm center}$. We need massless degrees of freedom (DOF) in the region $T_{\rm chiral} \leq T \leq T_{\rm center}$ to match the 't~Hooft anomaly.
The required DOF is very complicated, as explained in the main text.  \label{fig:phase1}}
\end{figure}

First, let us suppose that $T_{\rm chiral} < T_{\rm center}$ and see what is required in this case.
In this case, both the $\BZ_2^{\rm center}$ and $\SU(N_f)_L \times \SU(N_f)_R$ are unbroken at the intermediate temperature $T_{\rm chiral} \leq T \leq T_{\rm center}$.
Then we need to have some degrees of freedom which match the anomaly. See Figure~\ref{fig:phase1}.

For generic global anomalies there is a possibility that the anomaly is matched by topological quantum field theory 
(see e.g. \cite{Seiberg:2016rsg,Tachikawa:2016cha,Tachikawa:2016nmo,Wang:2017loc,Garcia-Etxebarria:2017crf,Tachikawa:2017gyf,Wang:2018edf,Lee:2018eqa,Hsieh:2018ifc} 
for the case of relativistic quantum field theory and more references therein for condensed matter contexts).
However, in the case of parity anomaly we need massless propagating degrees of freedom rather than topological quantum field theory 
which can be argued by a simple modification of the argument
given in section~5 of \cite{Garcia-Etxebarria:2017crf}. If the symmetry is unbroken, we need massless fermions for the anomaly matching.

First of all, the effective theory \eqref{eq:univ} of chiral symmetry restoration is not possible because the linear sigma model described by $\Phi$ does 
not have the parity anomaly. In the non-linear sigma model of Sec.~\ref{sec:chiral}, the topological term \eqref{eq:theta2} was crucial for the anomaly matching, and this term is possible
by the nontrivial target space topology of the sigma model. Therefore, the standard assumption \eqref{eq:univ} based on linear sigma model is excluded. 

To take into account the anomaly, we may introduce fermions $\lambda_L$ and $\overline{\lambda}_R$
which are in the fundamental representation of $\SU(N_f)_L$ and $\SU(N_f)_R$, respectively. Then we introduce a yukawa-like coupling between these fermions and $\Phi$ as
\beq
 \tr_f ( \Phi^\dagger \lambda_L  \lambda_R) + {\rm c.c.}.
\eeq
When the symmetry is broken by the VEV of $\Phi$, these fermions $\lambda_{L,R}$ get masses from this coupling.  
On the other hand, when the symmetry is restored, they become massless and may account for the parity anomaly.

However, the scenario above has the following flaw. It is reasonable to assume that the color degrees of freedom are confined in the three dimensional effective theory.
This is because the three dimensional $\SU(N_c)$ gauge fields $(A_C)_i~(i=1,2,3)$ after the reduction on $S^1$ is strongly coupled.\footnote{
 The assumption $T_{\rm chiral} < T_{\rm center}$ suggests that the color is completely confined even in the four dimensional sense $(A_C)_\mu~(\mu=1,2,3,4)$. }
 So we assume that the color is confined
 and $\lambda_L, \lambda_R$ are gauge singlets. 
 In Sec.~\ref{sec:QCD} we discussed that the symmetry (sub)group of the theory is \eqref{eq:glob} and in particular any gauge invariant quantity 
 must be invariant under the element $c'_2$ defined in \eqref{eq:div2p}.
Now suppose that the greatest common divisor ${\rm gcd}(N_c, N_f)$ between $N_c$ and $N_f$ is nontrivial, 
\beq
{\rm gcd}(N_c, N_f) \neq 1.
\eeq
Define $c_3=(c'_2)^{N_f/{\rm gcd}(N_c, N_f)}$ which is a nontrivial element when ${\rm gcd}(N_c, N_f) \neq 1$.
Then, regardless of the baryon charges of $\lambda_{L,R}$, these fermions transform nontrivially under $c_3$ because they are in the fundamental representation
of $\SU(N_f)_{L,R}$.
This is a contradiction. Therefore, if ${\rm gcd}(N_c, N_f) \neq 1$, the above scenario is excluded. 
This is in perfect agreement with the results of \cite{Shimizu:2017asf}. In that work, a very nontrivial anomaly was found under the condition ${\rm gcd}(N_c, N_f) \neq 1$
which cannot be matched by fermions alone. Here we got the same conclusion as in \cite{Shimizu:2017asf} by a different (but closely related) argument.

If ${\rm gcd}(N_c, N_f) =1$, the above argument cannot be directly applied. However, it is unlikely that 
the order of $T_{\rm chiral}$ and $T_{\rm center}$ changes depending on the precise value of ${\rm gcd}(N_c, N_f) $.
For example, consider the case $N_c=15$. Is it possible that $T_{\rm chiral} < T_{\rm center}$ for $N_f=4$ but $T_{\rm chiral} \geq T_{\rm center}$ for $N_f=3$ and $N_f=5$?
We do not see such kind of sensitivity on ${\rm gcd}(N_c, N_f) $ in the dynamics of exactly solvable theories such as supersymmetric QCD (see \cite{Intriligator:1995au} for a review).
Therefore, the above scenario is very unlikely even if ${\rm gcd}(N_c, N_f) =1$ at least for generic values of $(N_c, N_f)$.

There are more reasons to think that the above scenario is unlikely.
For generic values of $(N_f, N_c)$, the baryon charges of $\lambda_{L,R}$ must be taken to be some peculiar number to satisfy
the condition that they are invariant under $c'_2$. So we must demand that some baryons with complicated baryon charges become massless for the above scenario to be realized.

Moreover, if $N_c$ is even, all gauge invariant quantities are bosonic.
Then the fermions $\lambda_{L,R}$ cannot exist if they are gauge singlets. 

Therefore, we expect that for generic $(N_c, N_f)$ the above scenario is very unlikely.
It could still happen that for some special $N_f$ given by some critical value $N_f^*(N_c)$ which is a function of $N_c$, the above scenario in terms of the $\lambda_{L,R}$
is realized. For example, in supersymmetric QCD, it happens at $N_f^*(N_c)=N_c+1$ in which case $T_{\rm chiral}=0$ while the color is still confined.
We do not know such a critical value $N_f^*(N_c)$ in non-supersymmetric QCD, but notice that there is a difference from the case of supersymmetric QCD, because
it is impossible to have fermions if $N_c$ is even as mentioned above.

In the usual (non-supersymmetric) QCD, chiral symmetry breaking may happen at least for $N_f \leq N_c$. 
(It can of course happen for larger $N_f$, but the range $N_f \leq N_c$ might be safe from the experience of the real world $N_f \lsim N_c=3$ as well as the results in supersymmetric QCD.)
It is very likely that in this range of color and flavor numbers we have $T_{\rm chiral} \geq T_{\rm center}$.

There may be a way to avoid the above conclusion if we are willing to accept that some gauge degrees of freedom  (in the three dimensional sense $(A_C)_{i=1,2,3}$) are not confined.
For example, suppose that there is a $\U(1)$ gauge field in three dimensions, $(A_{\U(1)})_i~(i=1,2,3)$ which is not confined,
and that $\lambda_L, \lambda_R$ have charges $\pm 1$ under this $\U(1)$. Then these fermions are not gauge singlets anymore,
and the action of $c'_2$ can be compensated by $\U(1)$ gauge transformations. It is possible to see that such a scenario is consistent with the anomaly as follows.
Suppose that the Wilson line $W_{\rm quark}$ has a VEV of the form 
\beq
W_{\rm quark}=\diag(-1, +1,\cdots,+1). 
\eeq
Such a configuration is not a potential minimum at high temperature regime,
but for the sake of considering just the anomaly, we neglect that issue. 
The three dimensional $\SU(N_c)$ gauge group is spontaneously broken to $\SU(N_c-1) \times \U(1)$ by the above VEV of $W_{\rm quark}$, 
and after confinement of $\SU(N_c-1)$ we get a $\U(1)$ gauge field. The components of the quarks which are coupled to the eigenvalue $(-1)$ of $W_{\rm quark}$
give massless fermions coupled to the $\U(1)$ and they contribute to the anomaly as is clear from the derivation of the anomaly in Sec.~\ref{sec:QCD}.
Therefore, as far as the anomaly is concerned, it is possible that we have the fermions $(\lambda_L, \lambda_R)$ coupled to a $\U(1)$ gauge field. 
This (and other more complicated scenarios) remain as a logical possibility,
but this is a rather exotic scenario. Moreover, it is not clear whether the $\U(1)$ can really remain deconfined or not, because it is strongly coupled for small $N_f$.\footnote{
The $\U(1)$ gauge theory has the topological symmetry which is not present in the full theory, and it is natural to break them by introducing monopole
operators in three dimensions. Such monopole operators, if relevant in the RG sense, may force the $\U(1)$ theory to confine.
In that case, the fermions $\lambda_{L,R}$ are confined at long distance scales and the above scenario of deconfined $\U(1)$
is not realized. At least in the large $N_f$ expansion, the dimension $\Delta$ of monopole operators is quite small; for the minimal charge
monopole operators, the dimension is given by~\cite{Borokhov:2002ib,Pufu:2013vpa}
$\Delta=0.265 (2N_f)-0.0383+O(1/(2N_f))$ where we have taken into account the fact that there are $2N_f$ charged fermions $\lambda_L, \lambda_R$ in our case.
For example, for $N_f=2$ the monopole operators are very likely to be relevant and may trigger confinement of the $\U(1)$.}
Therefore, we assume that this does not happen and hence $T_{\rm chiral} \geq T_{\rm center}$ in the following.

Indeed, the inequality $T_{\rm chiral} \geq T_{\rm center}$ was seen~\cite{Aharony:2006da,Mandal:2011ws,Isono:2015uda} 
in a holographic model of QCD \cite{Sakai:2004cn}.
There is a one parameter family of such holographic models, 
and depending on the parameter we can realize both of the cases 
$T_{\rm chiral} = T_{\rm center}$ and $T_{\rm chiral} > T_{\rm center}$. The holographic QCD contains several degrees of freedom which were not present in the QCD,
and those degrees of freedom may be responsible for the difference. 

In fact, it is easy to realize the inequality $T_{\rm chiral} > T_{\rm center}$ without changing the anomaly by considering the following simple toy model.
Let us add to the QCD an elementary scalar field $S$ which is an $N_f \times N_f$ matrix and transforms under  $(g_L, g_R) \in \SU(N_f)_L \times \SU(N_f)_R$ as 
$g_L  S g_R^\dagger$. We may introduce a yukawa coupling
$
\tr_f (S^\dagger  \psi_L \psi_R)
$
between the elementary scalar $S$ and the quarks $\psi_L$ and $\psi_R$. By changing the potential $V(S)$ of the scalar field $S$,
we can break the chiral symmetry at an arbitrary scale $\vev{S} $. We may take the scale to be much larger than the dynamical scale $\Lambda$ of the gauge theory.
In this way we can realize the situation $T_{\rm chiral} >T_{\rm center}$. This is possible since we have two independent scales $\vev{S}$ and $\Lambda$.
The parameter mentioned above in the holographic QCD might be something like the VEV $\vev{S}$ which can be taken independent of $\Lambda$.

\begin{figure}
\centering
\includegraphics[width=0.7\textwidth]{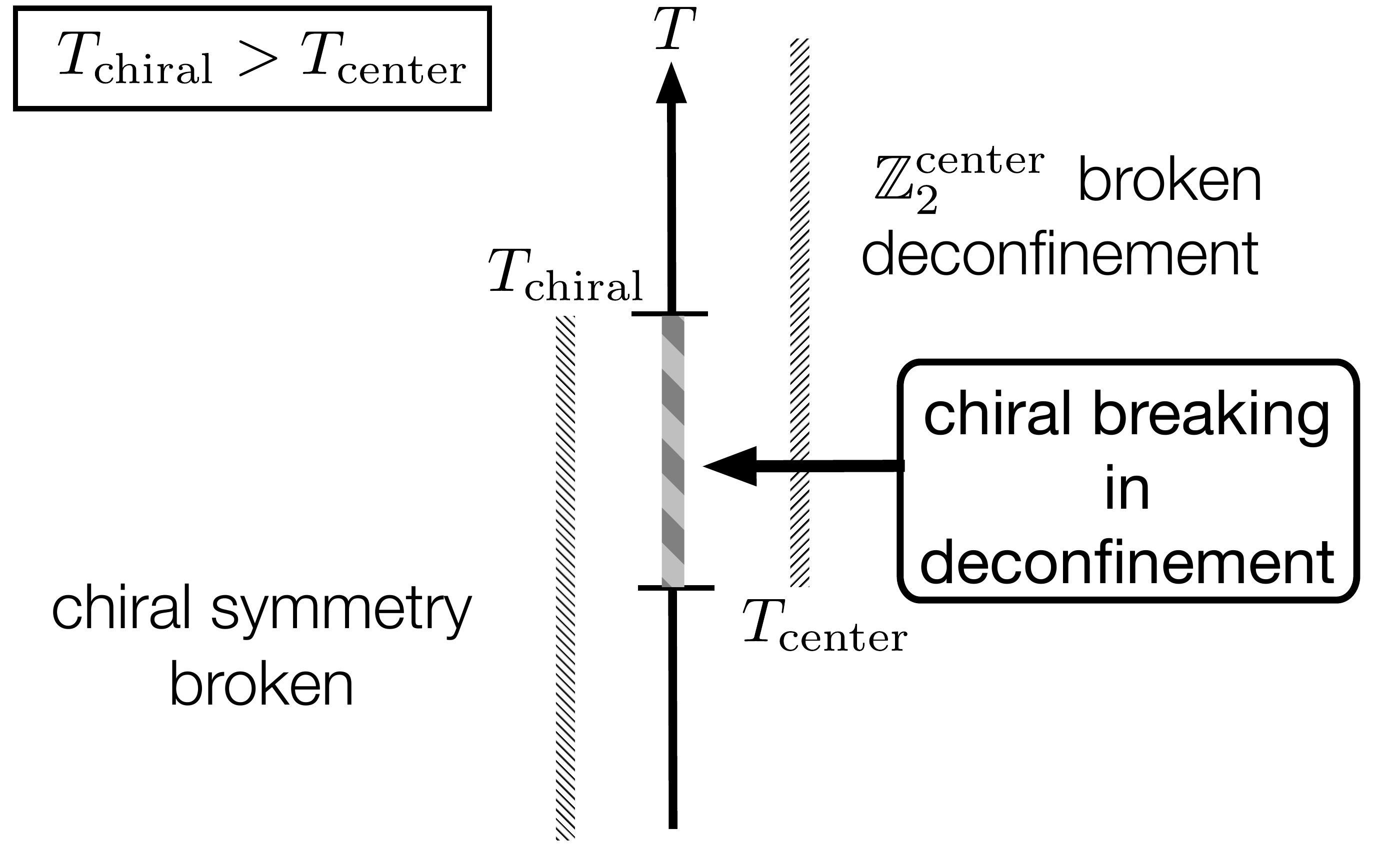}
\caption{The case  $T_{\rm chiral} > T_{\rm center}$. In the range $T_{\rm chiral} > T > T_{\rm center}$,
the chiral symmetry is broken even thougth the theory is in deconfinement phase.  \label{fig:phase2}}
\end{figure}

However, in the absence of such elementary scalar fields, it may be more natural to have the situation $T_{\rm chiral} = T_{\rm center}$ by the following reason.
Suppose contrary that we have $T_{\rm chiral} > T_{\rm center}$ as in Figure~\ref{fig:phase2}. Recall that $T_{\rm center}$ can be interpreted as the temperature of confinement/deconfinement
phase transition. Then, $T_{\rm chiral} > T_{\rm center}$ means that the chiral symmetry is broken in the deconfinement phase. 
If we introduce elementary scalar fields $S$ as above, it is easily realized. However, in the absence of such scalars, the chiral symmetry should be broken by quark condensates.
Intuitively, the quarks and anti-quarks are not so strongly bounded in the deconfinement phase and it is difficult to imagine that the quark condensation occurs in such a deconfinement phase. 
Although $T_{\rm chiral} > T_{\rm center}$ is allowed by the anomaly, it is rather exotic.
Therefore, the possibility $T_{\rm chiral} = T_{\rm center}$ is more natural in the QCD which contains only the dynamical scale $\Lambda$ as a parameter.
The fact that the holographic QCD realizes such a case at least for certain parameter range~\cite{Aharony:2006da,Mandal:2011ws,Isono:2015uda} is an evidence for this claim.
Therefore we may consider $T_{\rm chiral} = T_{\rm center}$ as the most natural possibility, and denote
this temperature as 
\beq
T_c:=T_{\rm chiral} = T_{\rm center}.\label{eq:singleT}
\eeq
This is the single critical temperature.

Finally, we need to ask whether the phase transition at $T_c$ is first order or second order. 
If it is second order, the situation is actually not so different from the case of $T_{\rm chiral} < T_{\rm center}$ which we have discussed above.
The reason is that at $T=T_c$ both of the symmetries $\BZ_2^{\rm center}$ and $\SU(N_f)_L \times \SU(N_f)_R$ are preserved 
because a second order phase transition is continuous (though not smooth) and these symmetries are unbroken at either $T<T_c$ or $T>T_c$.
Therefore, we need massless fermions like $\lambda_L$ and $\lambda_R$ which are coupled to a $\U(1)$ gauge field, or more exotic degrees of freedom.
Even worse, $T_{\rm chiral} = T_{\rm center}$ seems to be fine-tuning once we accept the appearance of $\lambda_{L,R}$,
because we can match the anomaly even if $T_{\rm chiral} < T_{\rm center}$ in that case. There seems to be no particular reason that $T_{\rm chiral} $ and $ T_{\rm center}$ coincides.
Therefore, we regard a second order transition at $T_c$ as unlikely for generic values of $(N_c, N_f)$.

\begin{figure}
\centering
\includegraphics[width=0.7\textwidth]{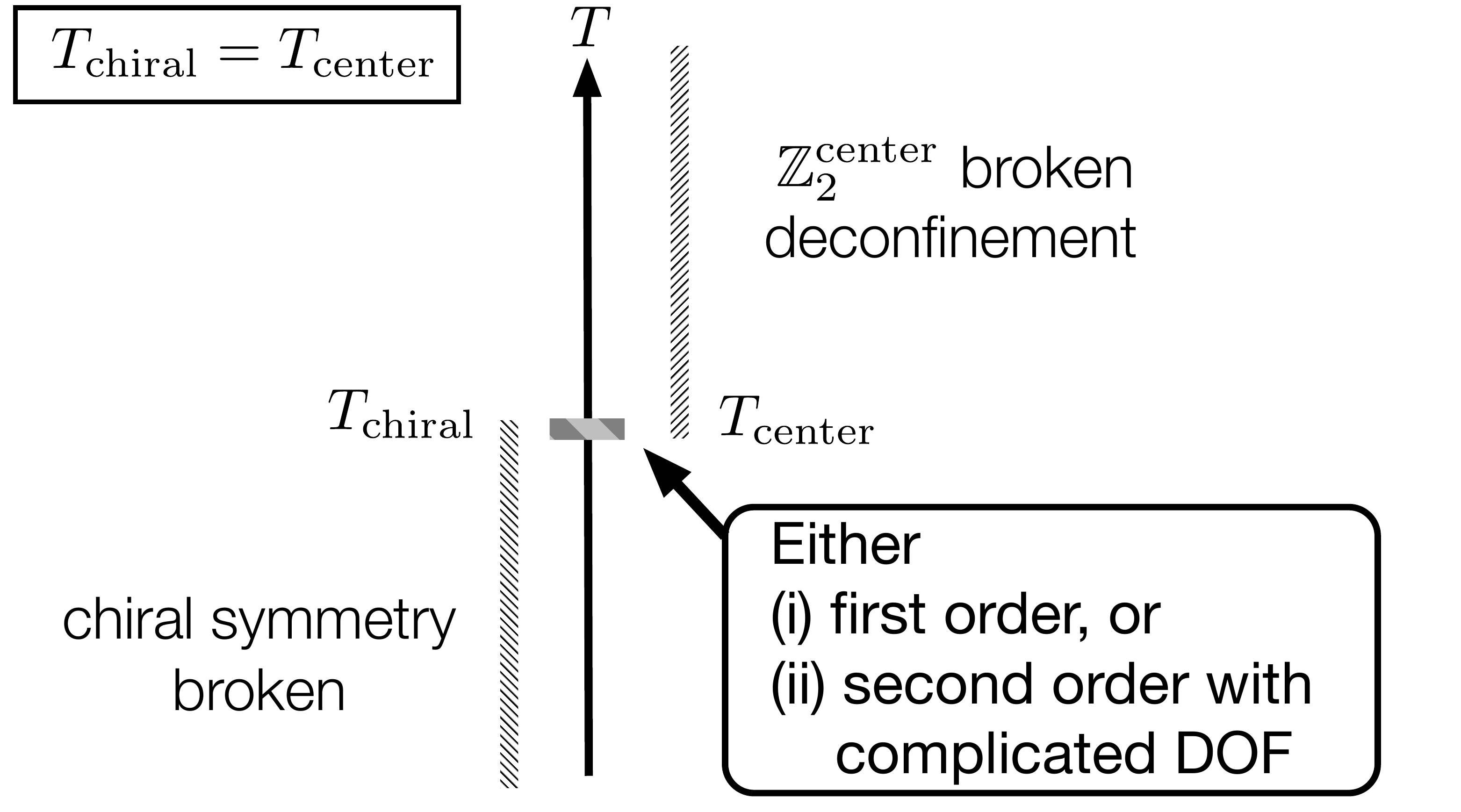}
\caption{The case  $T_{\rm chiral} = T_{\rm center}$. A first order phase transition is natural. If it is second order, some complicated massless DOF must appear at the critical temperature
to match the 't~Hooft anomaly.  \label{fig:phase3}}
\end{figure}

By the above arguments, we conclude that the most natural scenario may be a first order phase transition at the single critical temperature \eqref{eq:singleT} as in Figure~\ref{fig:phase3}.
This is what was discussed in \cite{Shimizu:2017asf}.
Of course, each of the above steps relies on some intuition about strong dynamics, and some exotic possibilities could be realized if such intuition is violated.
It would be very interesting to investigate these points further.

\subsubsection{$\mu_B=0$}
Let us next move to the case $\mu_B=0$. We argued above that a first order phase transition at a single critical temperature $T_c$
is the most natural possibility, at least for generic $(N_c, N_f)$ with $N_f \lsim N_c$. We assume that this is the case.
Can this conclusion change if we vary the value of $\mu_B$?

We have seen in the previous subsection that the dependence of the free energy on $\mu_B$ is given by \eqref{eq:count1} at high temperatures.
From this, we can also see that the entropy (defined formally as $S = - (\partial F)/(\partial T)$) behaves as
\beq
S \sim N_c^2 + N_f N_c + \frac{N_f}{N_c} \mu_B^2.
\eeq
The point is that $\mu_B$ only changes the entropy by an order of $N_f/N_c$.

On the other hand, in a first order phase transition, it is natural to think that the entropy changes by an order $N_c^2$ if $N_f \lsim N_c$.\footnote{
The reason that we use the entropy rather than the free energy is that the free energy $F$ is the same just before and after the phase transition (because
the true phase is obtained by finding the point which minimizes $F$, and the phase transition occurs when there are two points with the same minimum value of $F$).
For example, in the large $N_c$ limit of $\CN=4$ super-Yang-Mills on $S^3$, the free energy becomes $F \sim 1$ at the critical temperature $T_c$,
but the entropy is of order $N^2_c$ for $T=T_c+\epsilon$ and of order $1$ for $T=T_c -\epsilon$ for positive infinitesimal $\epsilon$. See \cite{Witten:1998zw}.}
For example, in the high and low temperature limits we have $S \sim c_{\rm high} N_c^2$ for $T \gg \Lambda$ from gluons and $S \sim c_{\rm low} N_f^2$ for $T \ll \Lambda$ from
Goldstone bosons with different numerical coefficients $c_{\rm high} \neq c_{\rm low}$. So let us assume that the entropy changes by an order of $N_c^2$.

Compared to the change of the entropy $\Delta S \sim N_c^2$ in the phase transition, the effect of the imaginary chemical potential $(N_f/N_c) \mu_B^2$ is sub-leading in the large $N$ expansion.
This is true even if the flavor number $N_f$ is comparable to $N_c$, $N_f \sim N_c$.
Therefore, it is natural to think that the imaginary chemical potential does not change the qualitative behavior of what we have found above, at least in the large $N_c$ limit.

We conclude that the first order phase transition is the most natural possibility even for the case $\mu_B=0$.
This argument is more solid for larger values of $N_c$. The value of interest in the real world, $N_c=3$, is a marginal value
because $N_c=2$ and $N_c \geq 3$ has different qualitative structure (e.g. whether the anti-fundamental representation of $\SU(N_c)$ is the same as the fundamental representation).
The qualitative success of large $N_c$ expansion in QCD phenomenology suggests that we should take large $N_c$ analysis seriously, as emphasized in \cite{Witten:1979kh}.

\subsubsection{Comparison with pure Yang-Mills at $\theta=\pi$ and $\theta=0$}
There is some analogy between QCD at finite temperature with $\mu_B=\pi$ and pure Yang-Mills at finite temperature with $\theta=\pi$.
In the case of pure Yang-Mills, the relevant symmetries (in three dimensions after compactification on the thermal circle $S^1$) are time-reversal symmetry $\sT$,
zero-form center symmetry $\BZ_{N_c}^{\rm center}$, and one-form center symmetry $\BZ_{N_c}^{\rm center(1)}$. There is a mixed anomaly among these three symmetries
which severely constrains the nature of thermal phase transition~\cite{Gaiotto:2017yup}.
We assume that the one-form center symmetry is always unbroken, and in the following, the zero-form center symmetry is just called the center symmetry.
We denote the critical temperatures associated to the symmetries $\sT$ and $\BZ_{N_c}^{\rm center}$
as $T_\sT$ and $T_{\rm center}$, respectively. The time-reversal is preserved at high temperature $T>T_\sT$ and broken at low temperature $T<T_\sT$.
The center symmetry is broken at high temperature $T>T_{\rm center}$ and preserved at low temperature $T<T_{\rm center}$.
Therefore, in the analogy with QCD, the time-reversal corresponds to $\SU(N_f)_L \times \SU(N_f)_R$, and $\BZ_{N_c}^{\rm center}$ corresponds to $\BZ_{2}^{\rm center}$.

For the pure Yang-Mills at $\theta=\pi$, let us repeat the argument we have given above for QCD.
First of all, suppose that $T_\sT< T_{\rm center}$. Then all the symmetries are preserved at $T_\sT \leq T \leq T_{\rm center}$, and we need complicated degrees of freedom
to match the anomaly. In fact, if we assume only the degrees of freedom which are naturally expected from the Yang-Mills, it is not possible to match the anomaly as discussed in detail in~\cite{Gaiotto:2017yup}.
Therefore, it is very likely that $T_\sT \geq T_{\rm center}$.

Next, let us ask whether $T_\sT > T_{\rm center}$ is natural or not. Recall that $T_{\rm center}$ is the temperature of confinement/deconfinement phase transition.
Then, if the inequality $T_\sT > T_{\rm center}$ is satisfied, that means that the time-reversal is broken in the deconfinement phase in the temperature range $T_{\rm center} < T < T_\sT$.
In pure Yang-Mills theory, one of the natural order parameters for the time-reversal at zero temperature is the parity odd glueball operator $G_A=\tr_c (F_{\mu\nu}F_{\rho\sigma})\epsilon^{\mu\nu\rho\sigma}$.
In fact, in the Euclidean path integral at zero temperature, $\vev{G_A} \propto -\frac{\partial }{\partial \theta} \log  Z = \frac{\partial }{\partial \theta} V(\theta)$
where $Z$ is the partition function and $V(\theta)$ is the vacuum energy as a function of $\theta$. At least in the large $N_c$ limit,
this potential behaves as $V(\theta) \propto \theta^2$~\cite{Witten:1979vv,Witten:1980sp,Witten:1998uka,Yonekura:2014oja,Yamazaki:2017ulc} and hence $\vev{G_A} \neq 0$.

However, there is a subtlety which was not present in QCD. Although $G_A$ is the most natural order parameter at zero temperature,
the 't~Hooft and dyonic line operators wrapped on the thermal circle $S^1$ are also good order parameters at finite temperature~\cite{Gaiotto:2017yup}.
To pursue the analogy with the QCD, we neglect this subtle issue and regard $G_A$ as the most relevant order parameter. 
Assuming this, the intuitive picture of the time-reversal breaking is that there is gluon condensation; from a pair of gluon fields $F_{\mu\nu}$ and $F_{\rho\sigma}$,
we make the operator corresponding to the bound state of the gluons as $G_A=\tr_c (F_{\mu\nu}F_{\rho\sigma})\epsilon^{\mu\nu\rho\sigma}$.
The time-reversal is broken by the condensation of this bound state. However, if the inequality $T_\sT > T_{\rm center}$ holds, that implies that we have the condensation of gluons
even in the deconfinement phase. This is counter-intuitive. Therefore, we regard this scenario to be unlikely, up to the subtlety discussed above.
Then $T_\sT = T_{\rm center}$ may be more natural and we assume it in the following.

Suppose that the phase transition at $T_c:=T_\sT = T_{\rm center}$ is second order. Then all the symmetries are preserved at $T=T_c$ and we need complicated degrees of freedom to match the
anomaly. As discussed for the case $T_\sT< T_{\rm center}$, we regard such complicated degrees of freedom to be unlikely. Therefore, we conclude that 
a first order phase transition at $T_c$ is the most natural possibility.

Finally let us change the value of $\theta$. By the standard large $N_c$ counting~\cite{Witten:1979vv,Witten:1980sp,Witten:1998uka},
the effect of $\theta$ to the partition function and hence the entropy is at most
\beq
S \sim N_c^2 + \theta^2.
\eeq
The actual effect is more suppressed, but that is not necessary for our purposes here. The point is that the dependence on $\theta$ is just a sub-leading effect in the large $N_c$ expansion.
On the other hand, in the first order deconfinement phase transition, it is natural to consider that the entropy changes by the order of $N_c^2$. 
Therefore, it seems unlikely that the value of $\theta$ affects the qualitative behavior of the first order phase transition if the large $N_c$ expansion is qualitatively good.

From the above considerations, we conclude that a first order phase transition is the most natural possibility even at $\theta=0$, if large $N_c$ expansion is qualitatively good.
This is what is strongly believed to be the case. For an explicit lattice simulation, see e.g. \cite{Panero:2009tv}, where first order transitions are seen for $N_c \geq 3$.
In holographic models, deconfinement phase transitions are always first order because they involve the change of the spacetime topology (see e.g \cite{Witten:1998zw,Aharony:2006da}).
Notice that the first order transition for large $N_c$ pure Yang-Mills at $\theta=0$ may not follow from the argument of universality.
Most naively, in the large $N_c$ limit the symmetry $\BZ_{N_c}^{\rm center}$ becomes effectively $\U(1)$, and the Polyakov loop operator $L$ is the order parameter of this symmetry,
so the universality class might seem to be that of $\O(2)$ Wilson-Fisher fixed point with the effective Lagrangian of the form $\CL \sim |\partial_i L|^2- (T-T_c)|L|^2+|L|^4$. 
This is not the case in large $N_c$ pure Yang-Mills.

Therefore we have ``successfully shown" the first order phase transition of large $N_c$ pure Yang-Mills theory from considerations of the anomaly and the large $N$ expansion.
Of course, as in the case of the discussions in QCD, each of the steps of the discussions was not rigorous, and in particular there was an additional subtlety in pure Yang-Mills
related to the order parameter of the breaking of $\sT$ as discussed above. But this ``success" might give some confidence that the discussions given above for the case of QCD is in the right direction.
Obviously more detailed studies from various directions would be desirable.

\subsection{Discussion}
The real QCD has two well-known small expansion parameters:
\begin{itemize}
\item The masses $m_q$ of the light quarks $q = u, d, (s)$.
\item $1/N_c$, where $N_c=3$.
\end{itemize}
At least at zero temperature, they are known to be good expansion parameters. Here, ``good" means that arguments based on the expansion about them 
explain many qualitative (even if not quantitative) properties of QCD dynamics at zero temperature.

In this paper, the large $N_c$ expansion has appeared in the discussions of how the thermodynamics depends on the imaginary baryon chemical potential $\mu_B$.
We mainly discussed the point $\mu_B=\pi$ to make the concept of confinement well-defined, but the effect of $\mu_B$ is highly suppressed in the large $N_c$ expansion.
Also, we focused our attention to $m_q=0$. 

In the applications of QCD phase transition to early cosmology, we are interested in the physical quark masses $m_q \neq 0$ and zero imaginary chemical potential $\mu_B=0$.
This physical situation is the same at the leading order of expansion to the situation we have studied in this paper, $m_q=0$ and $\mu_B=\pi$. 
The difference appears only at sub-leading orders in the expansion. 
The small parameter $m_q$ explicitly breaks the symmetry $\SU(N_f)_L \times \SU(N_f)_R$, and the small parameter $(\pi - \mu_B)/N_c$ explicitly breaks the symmetry $\BZ^{\rm center}_2$.
Now we can distinguish two cases of phase transition: first order or second order.
\begin{itemize}
\item If the phase transition is second order, then at the critical temperature, we have gapless degrees of freedom at $(m_q , \mu_B)=(0,\pi)$ which are associated to the symmetries 
$\SU(N_f)_L \times \SU(N_f)_R$ and $\BZ^{\rm center}_2$. These gapless degrees of freedom get masses by small explicit-breaking parameters $m_q$ or  $(\pi - \mu_B)/N_c$, and the phase transition becomes cross-over.
\item If the phase transition is first order at $(m_q , \mu_B)=(0,\pi)$, the order of the phase transition may not change as far as the parameters $m_q$ and $1/N_c$ are good small expansion parameters.
\end{itemize}
The 't~Hooft anomaly discussed in this paper constrains the nature of phase transition at $(m_q , \mu_B)=(0,\pi)$. There are still many logical possibilities allowed by the anomaly, some of which
are shown in Figures~\ref{fig:phase1}, \ref{fig:phase2}, and \ref{fig:phase3}. Based on some reasonable intuition about strong dynamics, we have argued that a first order phase transition in
Figure~\ref{fig:phase3} may be the
most natural scenario. If so, the phase transition at the physical point $m_q \neq 0$ and $\mu_B=0$ may also be first order if $m_q$ and $1/N_c$ are good expansion parameters.
In other words, if the QCD phase transition at the physical point is cross-over, some of the assumptions about QCD which are reasonable at zero temperature (e.g. expansion in terms of $m_q$ and $1/N_c$) must be violated
at finite temperatures. It is an important problem to settle this issue.
For the approximate concepts of chiral symmetry and confinement/deconfinement to be useful in QCD, the parameters $m_q$ and $1/N_c$ must be regarded as small.

\acknowledgments
The work of K.Y.~is supported by JSPS KAKENHI Grant-in-Aid (Wakate-B), No.17K14265.


\bibliographystyle{JHEP}
\bibliography{ref}

\end{document}